\documentclass[a4paper,11pt]{article}
\usepackage{jheppub}
\usepackage[utf8]{inputenc}
\usepackage{multirow}
\usepackage{longtable}
\usepackage{comment}
\usepackage{physics}
\usepackage{slashed}
\usepackage[dvipsnames]{xcolor}
\usepackage{ytableau}
\usepackage{url}

\usepackage{units}
\usepackage{amsthm}
\renewcommand{\arraystretch}{1.5}

\def \L {\mathcal{L}}

\def \vec#1{{\boldsymbol{#1}}}
\newcommand{\ov}[1]{\overline{#1}}
\newcommand{\wt}[1]{\widetilde{#1}}
\newcommand{\eps}{\epsilon}
\newcommand{\C}{\mathcal{C}}
\renewcommand{\O}{\mathcal{O}}
\newcommand{\T}{\mathcal{T}}

\newcommand{\B}{\mathcal{B}}
\newcommand{\fB}{\mathfrak{B}}
\newcommand{\und}[2]{\begin{subarray}{l} #1 \\ #2 \end{subarray}}
\renewcommand{\phi}{\varphi}

\DeclareMathOperator{\spa}{span}

\allowdisplaybreaks

\title{Basis for non-derivative baryon-number-violating operators}
\author[1]{Julian Heeck,%
\note{ORCID: \href{https://orcid.org/0000-0003-2653-5962}{0000-0003-2653-5962}}
}
\author[2]{Brandon B. Le%
\note{ORCID: \href{https://orcid.org/0009-0002-7354-9136}{0009-0002-7354-9136}}
}
\affiliation{Department of Physics, University of Virginia, 
Charlottesville, Virginia 22904-4714, USA}

\emailAdd{heeck@virginia.edu}
\emailAdd{sxh3qf@virginia.edu}

\arxivnumber{2604.25134}

\abstract{
We present minimal bases for all non-derivative baryon-number-violating operators in the Standard Model Effective Field Theory up to mass dimension 11, as well as for the symmetry-protected operators up to dimension 15. Compared to existing results, our bases generally contain fewer terms and simpler contractions, although we also identify select cases where a minimal basis is incompatible with simple structures.
}

\begin{document}
\maketitle
\flushbottom

\newpage

\section{Introduction}
\label{sec:introduction}

Baryon number violation (BNV) is one of the most sensitive probes of physics beyond the Standard Model (SM), especially if it leads to nucleon decays~\cite{FileviezPerez:2022ypk}. These processes are usually described through an effective field theory, i.e.~non-renormalizable operators, which provides a convenient systematic classification and ordering scheme by the operators' mass dimension $d$~\cite{Weinberg:1979sa,Weinberg:1980bf}. Renormalizable UV completions for these operators can of course also be constructed, see refs.~\cite{Heeck:2026dmh,Heeck:2026lhd} for an exhaustive review of this subject.
Unlike most other signatures in particle physics, BNV in the form of nuclear decays is sensitive to operators with mass dimension $d\gg 6$~\cite{Heeck:2019kgr}, seeing as it would induce decay rates of rough order $\Gamma\sim \Lambda_\text{QCD} (\Lambda_\text{QCD}^{d-4}/\Lambda^{d-4})^2$ just on dimensional grounds, where $\Lambda$ is the suppression scale of the dimension-$d$ effective operator and $\Lambda_\text{QCD} \sim \unit[0.3]{GeV}$ is a stand-in for all underlying QCD and nuclear scales. For new-physics scales as low as $\Lambda \sim \unit{TeV}$, this estimate gives rates $\Gamma \gtrsim (\unit[10^{32}]{yr})^{-1}$ for $d\leq 13$, well within reach of Super-Kamiokande, which, for clean signatures, is even sensitive to rates of order $(\unit[10^{34}]{yr})^{-1}$~\cite{Super-Kamiokande:2020wjk,Super-Kamiokande:2020tor}. A more careful calculation reveals that even operators at $d=14$ or $15$ could lead to testable signatures~\cite{Heeck:2019kgr,Heeck:2026dmh}. Upcoming detectors such as Hyper-Kamiokande, DUNE, and THEIA will explore even smaller rates or longer nuclear lifetimes, and hence be sensitive to even higher dimensional BNV operators.

Importantly, mass dimension $d$ is not the only distinguishing quantity of BNV operators, otherwise phenomenology would be entirely dominated by the $d=6$ operators~\cite{Weinberg:1979sa}, seeing as they are least suppressed by $\Lambda$. As shown already by Weinberg~\cite{Weinberg:1980bf}, operators of different mass dimension carry different amounts of baryon number $\Delta B$ and lepton number $\Delta L$~\cite{Heeck:2025btc}, with $\Delta B = 1$, $2$, and $3$ starting at $d=6$, $9$, and $15$, respectively. Since we do not know nature's selection rule for $(\Delta B,\Delta L)$, we do not know at which mass dimension $d$ BNV first appears. In other words, it is easy to impose baryon and lepton number symmetries that would forbid BNV up to some $d >6$~\cite{Weinberg:1980bf,Heeck:2019kgr}. A popular example is provided by operators with $(\Delta B , \Delta L ) = (2,0)$, which first appear at $d = 9$ and lead to neutron--antineutron oscillations that are under theoretical and experimental scrutiny~\cite{Proceedings:2020nzz,Broussard:2025opd}. Another example is provided by $(\Delta B,\Delta L) = (2,\pm2)$ operators that first appear at $d=12$~\cite{Arnold:2012sd,He:2021mrt} and lead to dinucleon decays into two leptons, searched for in Super-Kamiokande~\cite{Super-Kamiokande:2018apg}. Similarly, lepton flavor symmetries can also lead to selection rules that forbid BNV operators below a certain $d$; for example, ref.~\cite{Hambye:2017qix} studied symmetries that forbid BNV below $d=10$ but allow for the clean proton-decay channel $p\to \mu^+ \mu^+ e^-$, subsequently searched for in Super-Kamiokande~\cite{Super-Kamiokande:2020tor}.

Until we have experimentally determined how baryon and lepton number are broken in nature, we must remain agnostic and study BNV in all its varieties, including through very-high dimensional operators. This complicates comprehensive studies, given the exponential growth of BNV operator number with $d$~\cite{Heeck:2025btc}. Indeed, we have yet to explicitly write down all experimentally testable BNV operators! In this article, we aim to make progress in this direction by constructing a basis of non-derivative BNV operators up to $d=15$ that could subsequently be used for phenomenological studies and inform optimal search strategies.

We will work with the Standard Model Effective Field Theory (SMEFT), see refs.~\cite{Isidori:2023pyp,Aebischer:2025qhh} for recent reviews. The basic idea is simple: construct all possible Lorentz \& gauge-invariant operators by taking products of SM fields and covariant derivatives, then order them by mass dimension. The difficult part is to find the \emph{minimal} number of operators, or the minimal number of unknown Wilson coefficients at a given $d$, which define an operator basis. By now, such a basis of SMEFT operators has been constructed for 
$d=6$~\cite{Buchmuller:1985jz,Grzadkowski:2010es}, $d=7$~\cite{Lehman:2014jma,Liao:2016hru}, $d=8$~\cite{Li:2020gnx,Murphy:2020rsh}, and $d=9$~\cite{Li:2020xlh,Liao:2020jmn}. Ref.~\cite{Harlander:2023psl} has pushed this up to $d=12$ using the program \texttt{AutoEFT}~\cite{Harlander:2023ozs}, which builds on refs.~\cite{Fonseca:2019yya,Li:2020gnx,Li:2020xlh,Li:2022tec}.
Beyond that, Hilbert-series methods have been developed to \emph{count} the number of independent operators~\cite{Lehman:2015via,Lehman:2015coa,Henning:2015daa,Henning:2015alf}, providing an alternative to the more explicit traditional counting method developed in ref.~\cite{Fonseca:2019yya}, now integrated in the \texttt{Mathematica} package \texttt{Sym2Int}~\cite{Fonseca:2017lem}.
There are essentially two different kinds of bases in the literature, which notably differ in their treatment of repeated particles within an operator, say $QQQL$:
\begin{itemize}
\item Permutation-symmetry basis: the basis is built from irreducible representations of the relevant permutation group, $S_3$  in the $QQQL$ example due to the three repeated $Q$. The basis here contains three terms: one that is fully symmetric under permutations of $Q$, $\O^{\ytableausetup{boxsize=0.35em}\ydiagram{3}}_{QQQL}$, one fully antisymmetric, $\O^{\ytableausetup{boxsize=0.35em}\ydiagram{1,1,1}}_{QQQL}$, and one with mixed symmetry, $\O^{\ytableausetup{boxsize=0.35em}\ydiagram{2,1}}_{QQQL}$. The benefit of this basis is that it can be constructed algorithmically up to high mass dimension, currently to $d=12$ for the full SMEFT~\cite{Harlander:2023psl}, and that the counting of necessary Wilson coefficients is straightforward, seeing as they inherit their symmetry properties from the basis operators: the fully symmetric one comes with 30 coefficients $\lambda^{\ytableausetup{boxsize=0.35em}\ydiagram{3}}_{abcd}$ for three generations, the antisymmetric one with 3 coefficients $\lambda^{\ytableausetup{boxsize=0.35em}\ydiagram{1,1,1}}_{abcd}$ (purely because of the three lepton generations) and the mixed one with 24 coefficients $\lambda^{\ytableausetup{boxsize=0.35em}\ydiagram{2,1}}_{abcd}$, adding up to 57 in total.  In the Lagrangian, this would read $\sum_{\lambda  = {\ytableausetup{boxsize=0.35em}\ydiagram{3}},\,{\ytableausetup{boxsize=0.35em}\ydiagram{2,1}},\,{\ytableausetup{boxsize=0.35em}\ydiagram{1,1,1}}}\sum_{a,b,c,d}^3 y^{\lambda}_{abcd} \O^{\lambda}_{\und{QQQL}{abcd}}$, where $a,b,c,d$ run over the fermion generations. The disadvantage is the larger number of basis elements compared to the compact basis below. Using \texttt{AutoEFT}, we constructed this permutation-symmetry basis for non-derivative BNV SMEFT operators up to $d=15$~\cite{autoEFTbasis} to complement this article.
\item Minimal basis: in the more traditional approach of refs.~\cite{Buchmuller:1985jz,Grzadkowski:2010es,Lehman:2014jma,Murphy:2020rsh,Liao:2020jmn}, operator bases are constructed with the goal of having the smallest number of terms. $QQQL$ is here written as $\O^{(1)}_{QQQL}=\epsilon_{\alpha\beta\gamma} \epsilon_{jn}\epsilon_{km} (Q_a^{\alpha j} C Q_b^{\beta k}) (Q_c^{\gamma m} C L_d^n)$~\cite{Abbott:1980zj,Grzadkowski:2010es}, which is just one \emph{term},  included in the SMEFT Lagrangian as $\sum_{a,b,c,d}^3 y_{abcd}\O^{(1)}_{\und{QQQL}{abcd}} $. Operator and couplings satisfy the identity $\O^{(1)}_{\und{QQQL}{abcd}}+\O^{(1)}_{\und{QQQL}{bacd}} = \O^{(1)}_{\und{QQQL}{cbad}} + \O^{(1)}_{\und{QQQL}{cabd}}$, which ensures that the number of independent operators is again 57 and not $3^4=81$. One of the main drawbacks of this minimal basis is that it is difficult to construct, as is evident from the history of this example operator alone~\cite{Fonseca:2019yya,Fonseca:Mainz2023}.
\end{itemize}
The difference between these two kinds of bases can be substantial; for example, for the $d=12$ operator type $Q^6 L^2$, the permutation-symmetry basis consists of 74 terms~\cite{Harlander:2023psl}, while \texttt{Sym2Int} claims a minimal basis could get away with just 2~\cite{Fonseca:Mainz2023}. This can indeed be shown constructively, as one can form linear combinations of the permutation-basis operators to build the minimal basis -- essentially adding symmetric and antisymmetric operators to eliminate symmetries~\cite{Fonseca:Mainz2023} -- the downside being that this construction almost surely will not correspond to a human-readable contraction involving simple tensors such as Levi-Civita or Kronecker. Since the main benefit of a minimal basis is the compact form, one is then forced to \emph{guess} simple contractions of the involved fields and  check if they form a basis, a tedious method that is exactly how we obtained the results in this article.

Using \texttt{Sym2Int}'s ``number of terms'' as a guide and \texttt{GroupMath}'s ``Invariants''~\cite{Fonseca:2020vke} as the target space -- supplemented by the \texttt{grassmann} package~\cite{grassmann} whenever anticommutation relations are important -- we construct a compact basis for all non-derivative BNV operators with $d\leq 11$, as well as for the operators whose $(\Delta B,\Delta L)$ values first occur at $d=12$, 14, or 15, dubbed symmetry-protected operators in ref.~\cite{Heeck:2026dmh}. These are $(\Delta B,\Delta L) = (2,-2)$ and $(2,2)$ at $d=12$, $(1,5)$ at $d=14$, and $(1,-5)$, $(2,-4)$, $(2,4)$, and $(3,1)$ at $d=15$; no symmetry-protected operators appear at $d=13$. The restriction to non-derivative operators is in part to keep things manageable, and in part because derivative operators are generically suppressed in their contributions to nucleon decays compared to non-derivative operators~\cite{Weinberg:1980bf}.
We compare our results to existing compact bases for $d=6$, 7, 8, 9, as well as partial results for $d=12$, and find differences already starting at $d=8$: we are generally able to get away with fewer basis operators than other studies, matching \texttt{Sym2Int}'s numbers, and also use simpler contractions, eschewing vector and tensor operators almost entirely.
Another interesting finding is that, starting at $d=11$, there are cases for which the minimal compact operator basis cannot be realized using simple contractions, forcing us to either add one additional operator or accept somewhat ugly basis operators.

The remainder of this article is organized as follows. In section~\ref{sec:prelims}, we introduce our notation and conventions, notably formalizing what is meant by \emph{basis}. Section~\ref{sec:construction} explains our procedure for finding bases and section~\ref{sec:classification} provides an overview of the number of operator types by mass dimension, $(\Delta B,\Delta L)$, and field content. We compare our findings with the literature in section~\ref{sec:comparison}. Sections~\ref{sec:dim-6-ops} and~\ref{sec:dim-7-ops} list the basis operators for $d=6$ and $7$, respectively, taken from the literature for completeness. Section~\ref{sec:dim-8-ops} lists our basis of non-derivative BNV operators at $d=8$, section~\ref{sec:dim-9-ops} for $d=9$, section~\ref{sec:dim-10-ops} for $d=10$, section~\ref{sec:dim-11-ops} for $d=11$, and sections~\ref{sec:dim-12-ops}, \ref{sec:dim-14-ops}, \ref{sec:dim-15-ops} for the symmetry-protected operators at $d=12$, 14, and 15, respectively, all split by $(\Delta B,\Delta L)$ and sometimes further by field content. We conclude in section~\ref{sec:conclusions}. A dedicated discussion of the operator types for which no minimal basis consisting of nice terms exists is given in appendix~\ref{app:nonminimal}.

\section{Preliminaries}
\label{sec:prelims}

\begin{table}[tb]
\centering
{\renewcommand{\arraystretch}{1.1}
\begin{tabular}{c c c c} 
 \hline
 \multirow{2}{*}{field} & \multirow{2}{*}{chirality} & \multirow{2}{*}{generations} & $SU(3)_C\times SU(2)_L\times U(1)_Y$ \\ [-0.4ex]
  & & & representation \\
 \hline\hline
 $Q$ & left & 3 & $\left(\vec{3},\vec{2},\tfrac16\right)$\\
 $u$ & right & 3 & $\left(\vec{3},\vec{1},\tfrac23\right)$\\
 $d$ & right & 3 & $\left(\vec{3},\vec{1},-\tfrac13\right)$\\
 $L$ & left & 3 & $\left(\vec{1},\vec{2},-\tfrac12\right)$\\
 $e$ & right & 3 & $\left(\vec{1},\vec{1},-1\right)$\\
 $H$ & scalar & 1 & $\left(\vec{1},\vec{2},\tfrac12\right)$\\
 \hline
\end{tabular}
}
\caption{Standard Model fields and quantum numbers.}
\label{tab:fields}
\end{table}

The SMEFT is an effective field theory describing physics at energies above the electroweak scale and below a higher new-physics scale $\Lambda$. It is constructed from the SM field content, listed in table~\ref{tab:fields}, and inherits the SM gauge symmetry $SU(3)_C \times SU(2)_L \times U(1)_Y$. Its Lagrangian may be written schematically as
\begin{equation}
    \L_{\text{SMEFT}} = \L_{\text{SM}} + \sum_{d\geq 5}\sum_\T\sum_{\O\in\B_\T''}\frac{C}{\Lambda^{d-4}}\O \,,
    \label{eq:Lsmeft}
\end{equation}
where $\L_{\text{SM}}$ is the SM Lagrangian, $d$ is the mass dimension of the operator, the second sum is over all operator types $\T$ of dimension $d$, $\B_\T''$ is a basis of operators for the operator space $V_\T$, and $C$ is the Wilson coefficient corresponding to the operator $\O$.

Since the construction of SMEFT operators requires a substantial amount of notation, as evidenced by the previous sentence, we begin by fixing our conventions. First, we summarize the field, gauge, and index conventions used throughout. Then, we define the terminology and notation used to organize the operator expansion in eq.~\eqref{eq:Lsmeft}, distinguishing operator types, terms, and individual operators, together with the corresponding spaces in which linear relations are studied. These definitions provide the framework used in the following sections to construct minimal bases for BNV operators.

\subsection{Notation and conventions}

Our notation and conventions follow ref.~\cite{Liao:2020jmn}, with SM Lagrangian
\begin{equation}
\begin{split}
    \L_{\text{SM}} &= -\frac{1}{4}G_{\mu\nu}^AG^{A\mu\nu} - \frac{1}{4}W_{\mu\nu}^IW^{I\mu\nu} - \frac{1}{4}B_{\mu\nu}B^{\mu\nu} \\
    &\phantom{=}\ + (D_\mu H)^\dag(D^\mu H) - \lambda\qty(H^\dag H - \frac{1}{2}v^2)^2 \\
    &\phantom{=}\ + \sum_{\psi}\ov{\psi}i\slashed{D}\psi - \qty[\ov{L}Y_eeH + \ov{Q}Y_uu\wt{H} + \ov{Q}Y_ddH + \text{h.c.}].
\end{split}
\end{equation}
Here, $A$ and $I$ are $SU(3)_C$ and $SU(2)_L$ adjoint indices, respectively. $Y_e$, $Y_u$, and $Y_d$ are the Yukawa couplings, and $\wt{H} \equiv \eps H^*$, where $\eps$ is the Levi-Civita tensor for $SU(2)$. 
The Lorentz group is implemented as $SU(2)_l\times SU(2)_r$. Fermions are written as four-component Dirac spinors and generically represented as $\psi\in\{Q,u,d,L,e\}$, with $\ov{\psi} \equiv \psi^\dag\gamma^0$ denoting the Dirac adjoint. In classifying operators, we also denote left-handed fermions as $\ell\in\{Q,L\}$, right-handed fermions as $r\in\{u,d,e\}$, and scalar fields as $\varphi\in\{H,\wt{H}\}$. The gauge covariant derivative is defined as
\begin{equation}
    D_\mu = \partial_\mu - ig_3T^AG^A_\mu - ig_2T^IW^I_\mu - ig_1YB_\mu \,.
\end{equation}
Here, $g_{1,2,3}$ are the gauge couplings, and $T^A = \lambda^A/2$ and $T^I = \tau^I/2$ are the generators for $SU(3)_C$ and $SU(2)_L$, where $\lambda^A$ and $\tau^I$ are the Gell-Mann and Pauli matrices, respectively. The $U(1)_Y$ hypercharge $Y$ is related to the electric charge $Q$ by $Q = T^3 + Y$.

We now introduce the notation used in our operator contractions. We use superscript Greek letters $\alpha,\beta,\gamma,\delta,\kappa,\lambda,\rho,\sigma,\tau$ to denote field components in the fundamental representation of $SU(3)_C$ and superscript lowercase Latin letters $i,j,k,l,m,n,p,q$ to denote field components in the fundamental representation of $SU(2)_L$. Superscript uppercase Latin letters $I,J$ label the adjoint representation of $SU(2)_L$, and whenever they appear in a contraction, summation over them is implied. Finally, we use subscript lowercase Latin letters $a,b,c,d,e,f,g,h,r,s$ as flavor indices for the three fermion generations. For example, $Q^{i\alpha}_a$ denotes the component of the $a$th generation left-handed quark doublet with weak index $i$ and color index $\alpha$.

We further adopt the following conventions for explicit operator contractions. Parentheses $(\cdot)$ are used to group fields contracted into Lorentz scalars, as well as groups of Higgs fields. When a single Higgs doublet appears inside parentheses together with exactly one other $SU(2)_L$ doublet, the weak-index contraction is left implicit. For fermions, $C$ is the charge-conjugation matrix satisfying $C^T = C^\dag = -C$, $C^2 = -1$, and $\psi^{\C} \equiv C\ov{\psi}^{T}$ denotes the charge conjugated spinor, with opposite chirality of $\psi$. We also use the shorthand
\begin{equation}
    \psi_1 C \psi_2 \equiv \psi_1^{T} C \psi_2
\end{equation}
for fermion bilinears involving a charge conjugation operator. We write $H^\dag$ for the Hermitian conjugate Higgs doublet and $\wt{H}\equiv \eps H^*$ for the conjugate Higgs field.

We distinguish the invariant tensors of $SU(2)_L$ and $SU(3)_C$ explicitly. For $SU(2)_L$, we use the antisymmetric tensor $\eps_{ij}$, normalized by $\eps_{12}=+1$, and the Kronecker delta $\delta_{ij}$. The tensor $\eps_{ij}$ is used to contract two $SU(2)_L$ doublets antisymmetrically into a singlet, while $\delta_{ij}$ contracts a doublet with a conjugate doublet into a singlet. We also use the Pauli matrices $\tau^I_{ij}$, where $I$ is an $SU(2)_L$ adjoint index, to contract a doublet with a conjugate doublet into the adjoint representation, and the symmetric combination $(\eps\tau^I)_{ij}\equiv \eps_{ik}\tau^I_{kj}$ to contract two doublets into the adjoint representation. For $SU(3)_C$, we use the totally antisymmetric tensor $\eps_{\alpha\beta\gamma}$, normalized by $\eps_{123}=+1$, and the Kronecker delta $\delta_{\alpha\beta}$. The tensor $\eps_{\alpha\beta\gamma}$ contracts three color triplets into a singlet, while $\delta_{\alpha\beta}$ contracts a triplet with an antitriplet. Since we write all field indices as superscripts, both $\eps$ and $\delta$ are written with lower indices and used directly as invariant tensors in contractions. For color-octet contractions, we use the Gell-Mann matrices $(\lambda^A)^{\xi}{}_{\alpha}$, normalized by $\tr(\lambda^A\lambda^B) = 2\delta^{AB}$ and with repeated adjoint indices summed.

\subsection{Operator terminology}
\label{subsec:terminology}

We now establish our operator terminology, since the word operator is used in more than one sense in the literature. Throughout this paper, following ref.~\cite{Fonseca:2019yya}, an \emph{operator type} $\mathcal{T}$ specifies only the field content, restricted in this work to non-derivative operators, for example $\wt{H}\ov{e}Q^3L^2$. At this stage, no gauge or Lorentz contractions are implied beyond the choice of fields themselves.

For a fixed operator type $\mathcal{T}$ and a fixed set of flavor indices $\mathcal{F}$, we write $\mathcal{O}_{\und{\mathcal{T}}{\mathcal{F}}}^{(X)}$ for an \emph{operator}, by which we mean a single gauge- and Lorentz-invariant contraction of the fields of type $\mathcal{T}$, with all flavor indices specified. The superscript $(X)$ distinguishes the different possible invariant contractions for that same field content and flavor assignment. For example, if $\mathcal{T}=\wt{H}\ov{e}Q^3L^2$, then the operator
\begin{equation}
    \mathcal{O}_{\und{\wt{H}\ov{e}Q^3L^2}{abcdef}}^{(1)} = \eps_{\alpha\beta\gamma}\eps_{ij}\eps_{kl}\eps_{mn}(\ov{e}_aQ^{i\alpha}_b)(L^j_cCQ^{k\beta}_d)(L^l_eCQ^{m\gamma}_f)\wt{H}^n
\end{equation}
denotes one specific contraction with flavor labels $a,b,c,d,e,f$. With this convention, questions of equality, linear dependence, and linear independence are unambiguous, since they are asked directly in the vector space spanned by fully flavor-resolved operators.

It is often useful, however, to suppress the flavor information and keep only the common gauge and Lorentz structure. We denote such an object by $\mathcal{O}_{\mathcal{T}}^{(X)}$ and refer to it as a \emph{term}~\cite{Fonseca:2019yya}, where we can now more precisely call the superscript $(X)$ a term label. A term is therefore a flavor-unexpanded contraction pattern representing the whole family of operators obtained by assigning explicit flavor indices to that same structure. For example, the term
\begin{equation}
    \mathcal{O}_{\wt{H}\ov{e}Q^3L^2}^{(1)} = \eps_{\alpha\beta\gamma}\eps_{ij}\eps_{kl}\eps_{mn}(\ov{e}Q^{i\alpha})(L^jCQ^{k\beta})(L^lCQ^{m\gamma})\wt{H}^n
\end{equation}
stands for the class of all operators with this gauge and Lorentz contraction and all possible flavor assignments. We denote the corresponding set of operators by
\begin{equation}
    \left\{\mathcal{O}_{\und{\mathcal{T}}{\mathcal{F}}}^{(X)}\right\} = \left\{\mathcal{O}_{\und{\mathcal{T}}{11\cdots1}}^{(X)}, \mathcal{O}_{\und{\mathcal{T}}{11\cdots2}}^{(X)}, \hdots, \mathcal{O}_{\und{\mathcal{T}}{33\cdots3}}^{(X)}\right\}.
\end{equation}
In practice, we will usually abuse notation by displaying flavor labels in an explicit contraction even when speaking at the level of terms; whenever the flavor indices do not appear in the subscript of $\mathcal{O}_{\mathcal{T}}^{(X)}$, it should be understood that we mean the term, not a particular operator.

This leads naturally to several related spaces. For a fixed operator type $\mathcal{T}$, we define the \emph{operator space} $V_{\mathcal{T}}$ to be the vector space of all gauge- and Lorentz-invariant contractions with field content $\mathcal{T}$ and all flavor indices specified. Equivalently, $V_{\mathcal{T}}$ is the space spanned by all operators $\mathcal{O}_{\und{\mathcal{T}}{\mathcal{F}}}^{(X)}$ obtained by taking every possible singlet contraction, in the group-theoretic sense, of the fields in $\mathcal{T}$ for every flavor assignment $\mathcal{F}$. This is the ambient vector space in which linear relations, linear independence, and spanning are understood.

For a given term $\mathcal{O}_{\mathcal{T}}^{(X)}$, we then define the associated subspace
\begin{equation}
    V_{\mathcal{T}}^{(X)} = \mathrm{span}\left\{\mathcal{O}_{\und{\mathcal{T}}{\mathcal{F}}}^{(X)}\right\},
\end{equation}
namely the subspace of $V_{\mathcal{T}}$ spanned by all flavor realizations of that particular gauge and Lorentz contraction. The full operator space $V_{\mathcal{T}}$ is obtained by summing these subspaces over all distinct terms of type $\mathcal{T}$. By contrast, the set of terms itself is not a vector space in any natural sense; a term is best regarded as a label for one such family of operators, and linear dependence is defined only after passing to the corresponding elements of $V_{\mathcal{T}}$.

This distinction is important when discussing bases. Since linear independence is defined for fully specified operators, not for flavor-unexpanded terms by themselves, a basis of terms must be defined indirectly through the operators they generate. Accordingly, let
\begin{equation}
    \mathcal{B} = \left\{\mathcal{O}_{\mathcal{T}}^{(1)},\hdots,\mathcal{O}_{\mathcal{T}}^{(N)}\right\}
\end{equation}
be a set of terms of fixed operator type $\mathcal{T}$, and let
\begin{equation}
    \mathcal{B}' = \bigcup_{X=1}^N\left\{\mathcal{O}_{\und{\mathcal{T}}{\mathcal{F}}}^{(X)}\right\}
\end{equation}
be the set of all fully flavor-expanded operators arising from those terms. We say that $\mathcal{B}$ is a \emph{basis} for the operator type $\mathcal{T}$ if there exists a subset $\mathcal{B}''\subseteq \mathcal{B}'$ such that $\mathcal{B}''$ is linearly independent, spans the full operator space of type $\mathcal{T}$, and intersects the operator set $\left\{\mathcal{O}_{\und{\mathcal{T}}{\mathcal{F}}}^{(X)}\right\}$ coming from every term in $\mathcal{B}$. The last condition ensures that each term in $\mathcal B$ is represented in the chosen operator basis $\mathcal B''$. In this sense, a basis of terms is a collection of contractions from which a basis of the full operator space can be assembled, using at least one operator from every term. We call such a basis \emph{minimal} if no basis with fewer terms exists, and we denote a minimal basis for the operator type $\T$ as $\B_\T$. Then, the corresponding set $\B_\T''$ is a basis of operators for $V_\T$, so $|\B_\T''| = \dim V_\T$.

It is worth stressing that a basis of terms is not a basis in the usual linear-algebraic sense. If $\mathcal{B}_\T^1$ and $\mathcal{B}_\T^2$ are two bases of terms for the same operator type $\mathcal{T}$, and $\qty(\mathcal{B}_\T^1)''$ and $\qty(\mathcal{B}_\T^2)''$ are corresponding operator bases extracted from them, then
\begin{equation}
    \qty|\qty(\mathcal{B}_\T^1)''| = \qty|\qty(\mathcal{B}_\T^2)''| = \dim V_{\mathcal{T}} \,,
\end{equation}
since both are ordinary bases of the same operator space. By contrast, the numbers of terms in $\mathcal{B}_\T^1$ and $\mathcal{B}_\T^2$ need not agree:
\begin{equation}
    \qty|\mathcal{B}_\T^1| \neq \qty|\mathcal{B}_\T^2|
\end{equation}
in general. This is possible because a single term generally contains several linearly independent operators, so different choices of terms can organize the same operator space using different numbers of flavor-unexpanded structures. Only for minimal bases is the cardinality fixed by definition.

To study bases at the term level, it is useful to establish an intermediate point of view, in which we leave the flavor labels only formally expanded. Accordingly, for a fixed term $\mathcal{O}_{\mathcal{T}}^{(X)}$, we write $\mathcal{O}_{\und{\mathcal{T}}{\mathfrak{F}}}^{(X)}$, where $\mathfrak{F}$ denotes a formal pattern of flavor labels, written in the same placement convention as in the contraction itself. These labels are placeholders rather than fixed values in $\{1,2,3\}$. If the repeated fermion species in $\mathcal{T}$ occur with multiplicities $m_1,m_2,\ldots$, we let
\begin{equation}
    G_{\mathcal{T}} = S_{m_1}\times S_{m_2}\times\cdots,
\end{equation}
where each factor acts on the flavor labels belonging to one repeated field species. For example, for $\mathcal{T}=\wt{H}\ov{e}Q^3L^2$, we have $G_{\wt{H}\ov{e}Q^3L^2} = S_3\times S_2$, with $S_3$ acting on the three $Q$ flavor labels and $S_2$ acting on the two $L$ flavor labels.

For a fixed term $\mathcal{O}_{\mathcal{T}}^{(X)}$, we then define the corresponding \emph{flavor-permutation space} by
\begin{equation}
    W_{\mathcal{T}}^{(X)} = \mathrm{span}\left\{\mathcal{O}_{\und{\mathcal{T}}{\pi\cdot\mathfrak{F}}}^{(X)}: \pi\in G_{\mathcal{T}}\right\},
\end{equation}
where $\pi\cdot\mathfrak{F}$ denotes the formal flavor pattern obtained from $\mathfrak{F}$ by the action of $\pi$. We similarly define
\begin{equation}
    W_{\mathcal{T}}=\sum_X W_{\mathcal{T}}^{(X)},
\end{equation}
where the sum runs over all terms of type $\mathcal{T}$. Writing the full spanning set for $W_\T^{(X)}$ as
\begin{equation}
    \qty{\mathcal{O}_{\und{\mathcal{T}}{\mathfrak{F}}}^{(X)}} = \qty{\mathcal{O}_{\und{\mathcal{T}}{\pi\cdot\mathfrak{F}}}^{(X)}: \pi\in G_{\mathcal{T}}},
\end{equation}
given a set of terms $\B = \left\{\mathcal{O}_{\mathcal{T}}^{(1)},\hdots,\mathcal{O}_{\mathcal{T}}^{(N)}\right\}$, we can let
\begin{equation}
    \mathfrak{B}' = \bigcup_{X=1}^N\left\{\mathcal{O}_{\und{\mathcal{T}}{\mathfrak{F}}}^{(X)}\right\}
\end{equation}
be the set of all flavor-permuted operators corresponding to $\B$. Then, similarly, $\B$ is a basis of terms for $\T$ if there exists a subset $\mathfrak{B}''\subseteq\mathfrak{B}'$ forming a linearly independent spanning set for $W_\T$ that uses at least one operator from each term. For a minimal basis $\B_\T$ for $\T$, we denote the corresponding basis for $W_\T$ by $\mathfrak{B}_\T''$.

Thus, the flavor-permutation space $W_{\mathcal{T}}$ provides a convenient way to study the same basis problem at the level of terms. Indeed, each numerical flavor assignment gives an evaluation map from $W_{\mathcal{T}}$ to the fully flavor-expanded operator space $V_{\mathcal{T}}$, obtained by replacing the formal flavor labels by numerical values. Since $V_{\mathcal{T}}$ is generated by the images of these evaluation maps, any set of terms that spans $W_{\mathcal{T}}$ also spans $V_{\mathcal{T}}$ after full flavor expansion. However, we note that the minimum number of terms can decrease after the number of generations is fixed. For example, a term that is totally antisymmetric in four flavor slots of the same fermion species vanishes for any flavor assignment of three generations, even when it is nonzero in $W_{\mathcal T}$. For the operator types considered in this paper, we find that this subtlety only begins to have an effect at dimension 15.

\section{Construction of minimal bases}
\label{sec:construction}

Now that our terminology is established, we can formulate more precisely the goals of this work. Our primary aim is to construct explicit bases for non-derivative BNV SMEFT operator types that are both minimal and simple. Minimality is important because it allows the operator space to be described using the fewest terms possible. Since a term generally encodes more than a single direction in the fully flavored operator space, a smaller basis means that each chosen term captures as much of the space as possible, rather than distributing the same content across many similar expressions. This avoids unnecessary proliferation of similar terms and makes the overall operator space structure more transparent, providing a cleaner starting point for phenomenological applications such as matching and the study of ultraviolet completions.

At the same time, minimality alone is not the only consideration. In practice, one may encounter minimal bases whose elements are algebraically complicated, involving awkward linear combinations of contractions or Lorentz structures that obscure the underlying gauge structure. For many purposes, such a basis is less useful than a slightly larger but more transparent spanning set. A clean nonminimal basis can make symmetry properties, flavor permutations, and gauge contractions much easier to see, and can therefore be more valuable even when it contains redundancies. Our goal is more ambitious: whenever possible, we construct bases that are both minimal and composed of simple terms. These bases are not unique, and we do not claim that the choices presented here are in any absolute sense the best ones. Rather, our claim is that they are valid bases and, unless indicated otherwise, minimal ones, chosen to keep the contraction structure as simple and transparent as possible.

More specifically, the bases presented in this work are chosen to consist only of ``nice'' terms. By this, we mean that each basis element is written as one contraction of standard fermion bilinears and scalar fields, with gauge contractions written using only the invariant tensors $\eps_{\alpha\beta\gamma},\delta_{\alpha\beta},\eps_{ij},\delta_{ij},\tau^I_{ij},(\eps\tau^I)_{ij},(\lambda^A)^{\xi}{}_{\alpha}$, avoiding tensor currents whenever possible (although we allow these), and no basis elements being defined as explicit sums of distinct contraction patterns. Thus, each basis element is a single definite contraction written directly in terms of these standard invariant tensors. This makes the resulting bases easier to read, compare, and apply, while still retaining minimality.

With these goals in place, we now explain how the construction proceeds. First, we present a method for counting the number of terms in a minimal basis by exploiting permutation symmetries. Then, we describe the methodology used to construct our bases and to verify that they span the full operator space. Finally, we illustrate the procedure with the operator type $\wt{H}\ov{e}Q^3L^2$ as an explicit example.

\subsection{Counting of terms}
\label{subsec:term-counting}

By definition, the number of terms in a minimal basis is unambiguous. Here, we explain the algorithm for computing the number of terms $N_\T$ for a given operator type $\T$ by exploiting the permutation symmetries of its operators, following refs.~\cite{Fonseca:2019yya,Fonseca:Mainz2023}, corresponding to \texttt{Sym2Int}'s ``number of terms''. We assume that a permutation-symmetry basis for some operator type $\T$ has been found, either through \texttt{AutoEFT} or through \texttt{GroupMath}'s ``Invariants''. If $\T$ does not contain repeated fields, the permutation basis is already a minimal basis, otherwise the basis elements form irreducible representations $\lambda$ under permutations of repeated particles. In general, each representation comes with multiplicity $m_\lambda$, and the number of terms is given by
\begin{align}
N_\T = \left\lceil \max_\lambda \frac{m_\lambda}{\dim \lambda}  \right\rceil .
\label{eq:NT}
\end{align}
For example, $\T=QQQL$ breaks down into the aforementioned basis terms $\O^{\ytableausetup{boxsize=0.35em}\ydiagram{3}}_{QQQL}$, $\O^{\ytableausetup{boxsize=0.35em}\ydiagram{2,1}}_{QQQL}$, and $\O^{\ytableausetup{boxsize=0.35em}\ydiagram{1,1,1}}_{QQQL}$, each with multiplicity $m_\lambda = 1$. The $S_3$ representations have dimensions $\dim {\ytableausetup{boxsize=0.4em}\ydiagram{3}} =  \dim {\ytableausetup{boxsize=0.4em}\ydiagram{1,1,1}}=\tfrac12 \dim {\ytableausetup{boxsize=0.4em}\ydiagram{2,1}} = 1$, and so $N_\T = 1$, meaning one can take a linear combination of the three $\O^\lambda_{QQQL}$ to construct a minimal basis. One such linear combination would give the aforementioned basis term $\O^{(1)}_{QQQL}=\epsilon_{\alpha\beta\gamma} \epsilon_{jn}\epsilon_{km} (Q_a^{\alpha j} C Q_b^{\beta k}) (Q_c^{\gamma m} C L_d^n)$~\cite{Abbott:1980zj,Grzadkowski:2010es}.

For a slightly more complicated example, we take $\T=\wt{H}\ov{e}Q^3L^2$, with permutation symmetry $S_3(Q)\times S_2(L)$. The permutation-symmetry basis consists of 17 terms~\cite{Harlander:2023psl}:
\begin{align}
\ytableausetup{centertableaux}
3 \left(\ydiagram{3},\ydiagram{1,1}\right) + 4 \left(\ydiagram{2,1},\ydiagram{1,1}\right)+ 4 \left(\ydiagram{2,1},\ydiagram{2}\right)+ 2 \left(\ydiagram{1,1,1},\ydiagram{1,1}\right)+ 2 \left(\ydiagram{1,1,1},\ydiagram{2}\right)+ 2 \left(\ydiagram{3},\ydiagram{2}\right) ,
\end{align}
which gives
\begin{align}
N_{\wt{H}\ov{e}Q^3L^2} = \left\lceil \max\left\{ \frac{3}{1} ,\frac{4}{2} ,\frac{4}{2} ,\frac{2}{1} ,\frac{2}{1} ,\frac{2}{1}  \right\} \right\rceil = 3\,,
\end{align}
so a minimal compact basis has only 3 terms. These can be explicitly constructed by adding up the different $S_3\times S_2$ irreps to obtain symmetry-less terms:
\begin{align}
\O^{(1)}_\T &= a_1 \O^{\left(\ydiagram{3},\ydiagram{1,1}\right)}_{\T,1}+a_2 \O^{\left(\ydiagram{2,1},\ydiagram{1,1}\right)}_{\T,1}+a_3 \O^{\left(\ydiagram{2,1},\ydiagram{2}\right)}_{\T,1}+a_4 \O^{\left(\ydiagram{1,1,1},\ydiagram{1,1}\right)}_{\T,1}+a_5 \O^{\left(\ydiagram{1,1,1},\ydiagram{2}\right)}_{\T,1}+a_6 \O^{\left(\ydiagram{3},\ydiagram{2}\right)}_{\T,1}\,,\\
\O^{(2)}_\T &= b_1 \O^{\left(\ydiagram{3},\ydiagram{1,1}\right)}_{\T,2}+b_2 \O^{\left(\ydiagram{2,1},\ydiagram{1,1}\right)}_{\T,2}+b_3 \O^{\left(\ydiagram{2,1},\ydiagram{2}\right)}_{\T,2}+b_4 \O^{\left(\ydiagram{1,1,1},\ydiagram{1,1}\right)}_{\T,2}+b_5 \O^{\left(\ydiagram{1,1,1},\ydiagram{2}\right)}_{\T,2}+b_6 \O^{\left(\ydiagram{3},\ydiagram{2}\right)}_{\T,2}\,,\\
\O^{(3)}_\T &= c_1 \O^{\left(\ydiagram{3},\ydiagram{1,1}\right)}_{\T,3}\,,
\end{align}
with linear coefficients $a_j$, $b_j$, $c_j$. The terms $\O^{(1)}_\T$ and $\O^{(2)}_\T$ have no symmetries and thus come with $3^6=729$ Wilson coefficients each, while $\O^{(3)}_\T$ only has 90 coefficients given its manifest symmetries, adding up to 1548 coefficients/operators in total, in agreement with \texttt{Sym2Int} and ref.~\cite{Harlander:2023psl}.

Although the counting algorithm fixes the number of terms in a minimal basis, the basis elements it naturally produces are often unattractive from the standpoint of presentation. In particular, they almost always involve explicit linear combinations of several operators rather than single simple contractions. Because one of our goals is to present bases that are not only minimal when possible, but also simple and transparent, we therefore do not always insist on displaying a minimal basis for every operator type. For a few operator types, counting arguments show that a minimal basis consisting entirely of nice terms cannot exist. The operator types for which we do not present a minimal basis are listed in appendix~\ref{app:nonminimal}, together with some counting arguments that establish this impossibility. In cases where no such clear obstruction is known, however, it remains possible that a nice minimal basis does exist, but was not found in our construction.

\subsection{Methodology}
\label{subsec:methodology}

We now describe the general procedure used to construct the bases presented in this work. Although the final results are bases for the fully flavored operator spaces, it is technically simpler to carry out the construction in the flavor-permutation space $W_{\T}$ introduced in subsection~\ref{subsec:terminology}, rather than directly in $V_{\T}$. Roughly speaking, $W_{\T}$ keeps track of the linear relations implied by gauge and Lorentz contractions while treating the flavor slots formally, so that permutation properties among repeated fields can be analyzed cleanly before imposing the additional identifications present in the fully flavored space. We therefore carry out most spanning checks in $W_{\mathcal T}$, which guarantees spanning after the flavor labels take numerical values. For the dimension-15 types where the three-generation minimum is smaller (see sec.~\ref{subsec:2, 4, psi10}), we instead check spanning after imposing the three-generation flavor identities.

For a given operator type $\T$, the first step is to determine the dimension of $W_{\T}$. Equivalently, this is the number of linearly independent gauge- and Lorentz-singlet contractions that can be formed from the fields of type $\T$ in the flavor-permutation setting. To compute this, we treat each occurrence of a field as formally distinct, even when several fields have the same SM quantum numbers. In other words, repeated fields are not identified at this stage, which cleanly incorporates the separate generation slots and makes the counting a problem purely of gauge and Lorentz invariants. In practice, this can be computed using standard group-theoretic tools, for example, with the \texttt{Mathematica} package \texttt{GroupMath}, by computing the number of independent invariants under the Lorentz and gauge groups that can be formed from the fields in $\T$. This gives a definitive target dimension for the space we aim to span.

The second step is to determine the number of terms that can appear in a minimal basis, using the counting method described in subsection~\ref{subsec:term-counting}. Denoting this number by $N_{\T}$, the problem is then to construct a set of terms
\begin{equation}
    \B=\qty{\O_{\T}^{(1)},\dots,\O_{\T}^{(N_{\T})}}
\end{equation}
with $|\B|=N_{\T}$ such that the corresponding flavor-permuted operators span $W_{\T}$. Writing
\begin{equation}
    \mathfrak{B}'=\bigcup_{X=1}^{N_{\T}}
    \qty{\O_{\und{\T}{\pi\cdot\mathfrak F}}^{(X)}:\pi\in G_{\T}} \,,
\end{equation}
it is enough to exhibit a linearly independent subset $\mathfrak{B}''\subseteq\mathfrak{B}'$ with
\begin{equation}
    |\mathfrak{B}''|=\dim W_{\T} \,.
\end{equation}
Indeed, since every element of $\mathfrak{B}'$ lies in $W_{\T}$, we have
\begin{equation}
    \spa(\mathfrak{B}'')\subseteq \spa(\mathfrak{B}')\subseteq W_{\T} \,.
\end{equation}
But if $\mathfrak{B}''$ is linearly independent and has $\dim W_{\T}$ elements, then $\spa(\mathfrak{B}'')$ already has dimension $\dim W_{\T}$, and therefore
\begin{equation}
    \spa(\mathfrak{B}'')=\spa(\mathfrak{B}')=W_{\T} \,.
\end{equation}
Thus, once such a subset $\mathfrak{B}''$ is found, the set $\B$ is guaranteed to span the full flavor-permutation space.

The construction of $\B$ proceeds iteratively. We begin by guessing a first term $\O_{\T}^{(1)}$ that we expect to belong to the basis. As a guiding principle, we try to choose terms with as little permutation symmetry as possible among repeated fields. For example, we avoid contractions that are manifestly symmetric or antisymmetric under exchanging two identical field slots unless such symmetry is unavoidable. The reason is that permutation symmetries force linear relations among the operators in the associated set $\qty{\O_{\und{\T}{\mathfrak F}}^{(1)}} = \qty{\O_{\und{\T}{\pi\cdot\mathfrak F}}^{(1)}:\pi\in G_{\T}},$ thereby reducing the dimension of the subspace contributed by that term. A term with minimal permutation symmetry therefore tends to generate the largest possible subspace of $W_{\T}$ and is consequently the most efficient starting point for the basis construction.

Once a candidate term is chosen, we compute the explicit expressions for its flavor-permuted operators in \texttt{Mathematica} by carrying out the gauge and Lorentz contractions. In carrying out these contractions, we keep track of the Grassmann nature of the fermion fields, so permutations of fermionic factors are accompanied by the corresponding signs. This produces a concrete spanning set for $W_{\T}^{(1)}$. We then determine a linearly independent subset by solving linear systems among these operators in an explicit tensor realization. In practice, this means expanding all candidate operators in a common set of elementary tensor monomials and solving for linear relations among them. Our aim is usually to choose $\O_{\T}^{(1)}$ so that its associated flavor-permuted operators are already linearly independent, or as close to that as possible, since this maximizes the portion of $W_{\T}$ captured at the first step.

If the minimal number of terms $N_\T$ is greater than one, we then choose a second candidate term $\O_{\T}^{(2)}$. Here, the goal is not merely to contract the fields differently, but to do so in a way that contributes as many new independent directions as possible beyond those already spanned by the first term. In other words, we seek $\O_{\T}^{(2)}$ such that as many operators as possible in $\qty{\O^{(2)}_{\und{\T}{\mathfrak F}}}$ lie outside $W_{\T}^{(1)}$. After computing these operators explicitly, we again test which of them are independent of the previously chosen ones and enlarge our independent set accordingly. This process is then repeated term by term. At the $X$th stage, one chooses a term $\O_{\T}^{(X)}$ whose associated flavor-permuted operators are expected to contribute as many new independent directions as possible outside the span already obtained from $\O_{\T}^{(1)},\dots,\O_{\T}^{(X-1)}$.

The iteration terminates once we have constructed a set $\B$ with $|\B|=N_{\T}$ and found a subset $\mathfrak{B}''\subseteq\mathfrak{B}'$ consisting of $\dim W_{\T}$ linearly independent operators. At that point, the discussion above guarantees that $\B$ spans $W_{\T}$ and therefore furnishes a basis candidate with the correct number of terms. As a final consistency check, we verify that any other admissible operator of type $\T$ can be written as a linear combination of the operators in $\mathfrak{B}''$. This provides an explicit confirmation that no independent singlet contraction has been missed, and no mistakes were made.

Conceptually, the method combines three ingredients: a group-theoretic determination of $\dim W_{\T}$, a permutation-theoretic determination of the minimal number of terms $N_\T$, and an explicit constructive search for terms whose flavor orbits span the required space. The first two steps fix what must be achieved, while the third produces concrete bases realizing those requirements.

\subsection{\texorpdfstring{Illustrative example: $\wt{H}\ov{e}Q^3L^2$}{Illustrative example: HcecQ3L2}}

We now illustrate the general procedure of the previous subsection with the dimension-10 operator type $\T = \wt{H}\ov{e}Q^3L^2$, whose minimal basis is given in subsubsection~\ref{subsubsec:e*psi5H}. This example is useful because it is large enough to exhibit all of the main ingredients of the construction, but still small enough that the resulting basis and the associated linear relations can be written explicitly without too much trouble. Also, to our knowledge, a minimal basis for this operator type has not previously been given in the literature. As mentioned in subsection~\ref{subsec:term-counting}, the 17-element permutation-symmetry basis for this term has been constructed in ref.~\cite{Harlander:2023psl}.

The first step is to determine the target dimension of the flavor-permutation space. Treating the repeated fields as formally distinct and counting the independent Lorentz- and gauge-invariant singlets that can be formed from the fields in $\T = \wt{H}\ov{e}Q^3L^2$, we use \texttt{GroupMath} to find $\dim W_{\wt{H}\ov{e}Q^3L^2} = 25$. This is the dimension that any successful construction must reproduce. Independently, the term-counting method described earlier, implemented by \texttt{Sym2Int}, shows that a minimal basis for this operator type must contain 3 terms. Thus, before doing any explicit linear-algebra check, we already know that we are aiming for a set of 3 terms whose flavor orbits together span a 25-dimensional space.

We choose the three candidate terms
\begin{equation}
    \B_{\wt{H}\ov{e}Q^3L^2} = \qty{\O^{(1)}_{\wt{H}\ov{e}Q^3L^2}, \O^{(2)}_{\wt{H}\ov{e}Q^3L^2}, \O^{(3)}_{\wt{H}\ov{e}Q^3L^2}}
\end{equation}
given explicitly below:
\begin{align}
    \O^{(1)}_{\wt{H}\ov{e}Q^3L^2} &= \eps_{\alpha\beta\gamma}\eps_{ij}\eps_{kl}\eps_{mn}(\ov{e}_aQ^{i\alpha}_b)(L^j_cCQ^{k\beta}_d)(L^l_eCQ^{m\gamma}_f)\wt{H}^n  \,,\\
    \O^{(2)}_{\wt{H}\ov{e}Q^3L^2} &= \eps_{\alpha\beta\gamma}\eps_{ik}\eps_{jm}\eps_{ln}(\ov{e}_aQ^{i\alpha}_b)(L^j_cCQ^{k\beta}_d)(L^l_eCQ^{m\gamma}_f)\wt{H}^n  \,,\\
    \O^{(3)}_{\wt{H}\ov{e}Q^3L^2} &= \eps_{\alpha\beta\gamma}\eps_{il}\eps_{jk}(H^\dag\ov{e}_aL_b)(L^i_cCQ^{j\alpha}_d)(Q^{k\beta}_eCQ^{l\gamma}_f) \,.
\end{align}
These are three inequivalent contraction patterns that are expected to generate large subspaces under flavor permutation. Therefore, they are candidates for a minimal spanning set at the level of terms, and the task is to verify that they indeed suffice.

The relevant flavor-permutation group is
\begin{equation}
    G_{\wt{H}\ov{e}Q^3L^2} = S_3\times S_2  \,.
\end{equation}
For $\O^{(1)}$ and $\O^{(2)}$, the $S_3$ factor permutes the three $Q$ flavor slots $b,d,f$ and the $S_2$ factor permutes the two $L$ flavor slots $c,e$. For $\O^{(3)}$, because the fields are grouped differently, the same abstract group acts on different flavor slots: the $S_3$ acts on $d,e,f$ and the $S_2$ acts on $b,c$. As discussed in subsection~\ref{subsec:terminology}, the permutation group is fixed by the repeated fields in the operator type, while the slots on which it acts depend on the term.

Applying all flavor permutations to the three terms produces the full set of flavor-permuted operators associated with this candidate basis. We denote by $\fB'_{\wt{H}\ov{e}Q^3L^2}$ the union of these three orbits. Since each of $\O^{(1)}_{\wt{H}\ov{e}Q^3L^2}$, $\O^{(2)}_{\wt{H}\ov{e}Q^3L^2}$, and $\O^{(3)}_{\wt{H}\ov{e}Q^3L^2}$ has a full $S_3\times S_2$ orbit of size 12, the total number of orbit elements is $\qty|\fB_{\wt{H}\ov{e}Q^3L^2}'| = 12\times 3 = 36$.
Of course, these 36 operators cannot all be linearly independent, since we already know that the ambient space $W_{\T}$ has dimension 25. Thus, the problem is to extract from $\fB'$ a linearly independent subset of size 25.

A convenient choice of such a subset is the set $\fB''_{\wt{H}\ov{e}Q^3L^2}$ displayed below:
\begin{equation}
\begin{split}
    \fB_{\wt{H}\ov{e}Q^3L^2}'' &= \Bigg\{\O^{(1)}_{\und{\wt{H}\ov{e}Q^3L^2}{abcdef}}, \O^{(1)}_{\und{\wt{H}\ov{e}Q^3L^2}{abcfed}}, \O^{(1)}_{\und{\wt{H}\ov{e}Q^3L^2}{adcbef}}, \O^{(1)}_{\und{\wt{H}\ov{e}Q^3L^2}{adcfeb}}, \O^{(1)}_{\und{\wt{H}\ov{e}Q^3L^2}{afcbed}}, \O^{(1)}_{\und{\wt{H}\ov{e}Q^3L^2}{afcdeb}}, \\
    &\phantom{=}\quad\ \O^{(1)}_{\und{\wt{H}\ov{e}Q^3L^2}{abedcf}}, \O^{(1)}_{\und{\wt{H}\ov{e}Q^3L^2}{abefcd}}, \O^{(1)}_{\und{\wt{H}\ov{e}Q^3L^2}{adebcf}}, \O^{(1)}_{\und{\wt{H}\ov{e}Q^3L^2}{adefcb}}, \O^{(1)}_{\und{\wt{H}\ov{e}Q^3L^2}{afebcd}}, \O^{(1)}_{\und{\wt{H}\ov{e}Q^3L^2}{afedcb}}, \\
    &\phantom{=}\quad\ \O^{(2)}_{\und{\wt{H}\ov{e}Q^3L^2}{abcdef}}, \O^{(2)}_{\und{\wt{H}\ov{e}Q^3L^2}{abcfed}}, \O^{(2)}_{\und{\wt{H}\ov{e}Q^3L^2}{adcbef}}, \O^{(2)}_{\und{\wt{H}\ov{e}Q^3L^2}{adcfeb}}, \O^{(2)}_{\und{\wt{H}\ov{e}Q^3L^2}{afcbed}}, \O^{(2)}_{\und{\wt{H}\ov{e}Q^3L^2}{afcdeb}}, \\
    &\phantom{=}\quad\ \O^{(3)}_{\und{\wt{H}\ov{e}Q^3L^2}{abcdef}},\O^{(3)}_{\und{\wt{H}\ov{e}Q^3L^2}{abcdfe}},\O^{(3)}_{\und{\wt{H}\ov{e}Q^3L^2}{abcedf}},\O^{(3)}_{\und{\wt{H}\ov{e}Q^3L^2}{abcefd}},\O^{(3)}_{\und{\wt{H}\ov{e}Q^3L^2}{acbdef}},\O^{(3)}_{\und{\wt{H}\ov{e}Q^3L^2}{acbdfe}},\O^{(3)}_{\und{\wt{H}\ov{e}Q^3L^2}{acbedf}}\Bigg\} .
\end{split}
\end{equation}
Its size is $\qty|\fB_{\wt{H}\ov{e}Q^3L^2}''| = \dim W_{\wt{H}\ov{e}Q^3L^2} = 25$, so if this set is linearly independent, it is automatically a basis of $W_{\T}$. In practice, the independence check is carried out exactly as described in subsection~\ref{subsec:methodology}: one expands all operators into explicit component products of the fields, keeping the signs induced by the Grassmann nature of the fermion fields, and then solves for linear relations among the resulting expressions.

Comparing $\fB''$ with the full orbit set $\fB'$, we find the $36-25=11$ operators that are omitted:
\begin{equation}
\begin{split}
    \fB_{\wt{H}\ov{e}Q^3L^2}'\setminus \fB_{\wt{H}\ov{e}Q^3L^2}'' &= \Bigg\{\O^{(2)}_{\und{\wt{H}\ov{e}Q^3L^2}{abedcf}}, \O^{(2)}_{\und{\wt{H}\ov{e}Q^3L^2}{abefcd}}, \O^{(2)}_{\und{\wt{H}\ov{e}Q^3L^2}{adebcf}}, \O^{(2)}_{\und{\wt{H}\ov{e}Q^3L^2}{adefcb}}, \O^{(2)}_{\und{\wt{H}\ov{e}Q^3L^2}{afebcd}}, \O^{(2)}_{\und{\wt{H}\ov{e}Q^3L^2}{afedcb}}, \\
    &\phantom{=}\quad\ \O^{(3)}_{\und{\wt{H}\ov{e}Q^3L^2}{abcfde}},\O^{(3)}_{\und{\wt{H}\ov{e}Q^3L^2}{abcfed}},\O^{(3)}_{\und{\wt{H}\ov{e}Q^3L^2}{acbefd}},\O^{(3)}_{\und{\wt{H}\ov{e}Q^3L^2}{acbfde}},\O^{(3)}_{\und{\wt{H}\ov{e}Q^3L^2}{acbfed}}\Bigg\} .
\end{split}
\end{equation}
The fact that $\fB_{\wt{H}\ov{e}Q^3L^2}''$ is linearly independent and contains exactly $\dim W_{\wt{H}\ov{e}Q^3L^2}$ operators is sufficient to show that it is a basis of $W_{\wt{H}\ov{e}Q^3L^2}$, and hence that $\B_{\wt{H}\ov{e}Q^3L^2}$ is a basis for the operator type $\wt{H}\ov{e}Q^3L^2$. Nevertheless, as a check, we can verify explicitly that $\fB_{\wt{H}\ov{e}Q^3L^2}''$ spans $W_{\wt{H}\ov{e}Q^3L^2}$ by expressing each of the remaining 11 operators as a linear combination of the 25 elements of $\fB_{\wt{H}\ov{e}Q^3L^2}''$:
\begin{align}
    \O^{(2)}_{\und{\wt{H}\ov{e}Q^3L^2}{abedcf}} &= + \O^{(1)}_{\und{\wt{H}\ov{e}Q^3L^2}{abcfed}} + \O^{(1)}_{\und{\wt{H}\ov{e}Q^3L^2}{abedcf}} - \O^{(2)}_{\und{\wt{H}\ov{e}Q^3L^2}{abcfed}} \,,\\[2ex]
    \O^{(2)}_{\und{\wt{H}\ov{e}Q^3L^2}{abefcd}} &= + \O^{(1)}_{\und{\wt{H}\ov{e}Q^3L^2}{abcdef}} + \O^{(1)}_{\und{\wt{H}\ov{e}Q^3L^2}{abefcd}} - \O^{(2)}_{\und{\wt{H}\ov{e}Q^3L^2}{abcdef}}  \,,\\[2ex]
    \O^{(2)}_{\und{\wt{H}\ov{e}Q^3L^2}{adebcf}} &= + \O^{(1)}_{\und{\wt{H}\ov{e}Q^3L^2}{adcfeb}} + \O^{(1)}_{\und{\wt{H}\ov{e}Q^3L^2}{adebcf}} - \O^{(2)}_{\und{\wt{H}\ov{e}Q^3L^2}{adcfeb}}  \,,\\[2ex]
    \O^{(2)}_{\und{\wt{H}\ov{e}Q^3L^2}{adefcb}} &= + \O^{(1)}_{\und{\wt{H}\ov{e}Q^3L^2}{adcbef}} + \O^{(1)}_{\und{\wt{H}\ov{e}Q^3L^2}{adefcb}} - \O^{(2)}_{\und{\wt{H}\ov{e}Q^3L^2}{adcbef}}  \,,\\[2ex]
    \O^{(2)}_{\und{\wt{H}\ov{e}Q^3L^2}{afebcd}} &= + \O^{(1)}_{\und{\wt{H}\ov{e}Q^3L^2}{afcdeb}} + \O^{(1)}_{\und{\wt{H}\ov{e}Q^3L^2}{afebcd}} - \O^{(2)}_{\und{\wt{H}\ov{e}Q^3L^2}{afcdeb}}  \,,\\[2ex]
    \O^{(2)}_{\und{\wt{H}\ov{e}Q^3L^2}{afedcb}} &= + \O^{(1)}_{\und{\wt{H}\ov{e}Q^3L^2}{afcbed}} + \O^{(1)}_{\und{\wt{H}\ov{e}Q^3L^2}{afedcb}} - \O^{(2)}_{\und{\wt{H}\ov{e}Q^3L^2}{afcbed}}  \,,\\[2ex]
    \O^{(3)}_{\und{\wt{H}\ov{e}Q^3L^2}{abcfde}} &= - \O^{(3)}_{\und{\wt{H}\ov{e}Q^3L^2}{abcdfe}} + \O^{(3)}_{\und{\wt{H}\ov{e}Q^3L^2}{abcedf}} + \O^{(3)}_{\und{\wt{H}\ov{e}Q^3L^2}{abcefd}}  \,,\\[2ex]
    \O^{(3)}_{\und{\wt{H}\ov{e}Q^3L^2}{abcfed}} &= + \O^{(3)}_{\und{\wt{H}\ov{e}Q^3L^2}{abcdef}} + \O^{(3)}_{\und{\wt{H}\ov{e}Q^3L^2}{abcdfe}} - \O^{(3)}_{\und{\wt{H}\ov{e}Q^3L^2}{abcefd}}  \,,\\[2ex]
    \begin{split}
        \O^{(3)}_{\und{\wt{H}\ov{e}Q^3L^2}{acbefd}} &= -\, \O^{(1)}_{\und{\wt{H}\ov{e}Q^3L^2}{abedcf}} + \O^{(1)}_{\und{\wt{H}\ov{e}Q^3L^2}{abefcd}} + \O^{(1)}_{\und{\wt{H}\ov{e}Q^3L^2}{adebcf}} - \O^{(1)}_{\und{\wt{H}\ov{e}Q^3L^2}{adefcb}} - \O^{(1)}_{\und{\wt{H}\ov{e}Q^3L^2}{afebcd}} + \O^{(1)}_{\und{\wt{H}\ov{e}Q^3L^2}{afedcb}} \\
        &\phantom{=}\ - \O^{(2)}_{\und{\wt{H}\ov{e}Q^3L^2}{abcdef}} + \O^{(2)}_{\und{\wt{H}\ov{e}Q^3L^2}{abcfed}} + \O^{(2)}_{\und{\wt{H}\ov{e}Q^3L^2}{adcbef}} - \O^{(2)}_{\und{\wt{H}\ov{e}Q^3L^2}{adcfeb}} - \O^{(2)}_{\und{\wt{H}\ov{e}Q^3L^2}{afcbed}} + \O^{(2)}_{\und{\wt{H}\ov{e}Q^3L^2}{afcdeb}} \\
        &\phantom{=}\ + \O^{(3)}_{\und{\wt{H}\ov{e}Q^3L^2}{abcdfe}} - \O^{(3)}_{\und{\wt{H}\ov{e}Q^3L^2}{abcefd}} + \O^{(3)}_{\und{\wt{H}\ov{e}Q^3L^2}{acbdfe}}  \,,
    \end{split}  \\[2ex]
    \begin{split}
        \O^{(3)}_{\und{\wt{H}\ov{e}Q^3L^2}{acbfde}} &= -\, \O^{(1)}_{\und{\wt{H}\ov{e}Q^3L^2}{abedcf}} + \O^{(1)}_{\und{\wt{H}\ov{e}Q^3L^2}{abefcd}} + \O^{(1)}_{\und{\wt{H}\ov{e}Q^3L^2}{adebcf}} - \O^{(1)}_{\und{\wt{H}\ov{e}Q^3L^2}{adefcb}} - \O^{(1)}_{\und{\wt{H}\ov{e}Q^3L^2}{afebcd}} + \O^{(1)}_{\und{\wt{H}\ov{e}Q^3L^2}{afedcb}} \\
        &\phantom{=}\ - \O^{(2)}_{\und{\wt{H}\ov{e}Q^3L^2}{abcdef}} + \O^{(2)}_{\und{\wt{H}\ov{e}Q^3L^2}{abcfed}} + \O^{(2)}_{\und{\wt{H}\ov{e}Q^3L^2}{adcbef}} - \O^{(2)}_{\und{\wt{H}\ov{e}Q^3L^2}{adcfeb}} - \O^{(2)}_{\und{\wt{H}\ov{e}Q^3L^2}{afcbed}} + \O^{(2)}_{\und{\wt{H}\ov{e}Q^3L^2}{afcdeb}} \\
        &\phantom{=}\ + \O^{(3)}_{\und{\wt{H}\ov{e}Q^3L^2}{abcdfe}} - \O^{(3)}_{\und{\wt{H}\ov{e}Q^3L^2}{abcefd}} + \O^{(3)}_{\und{\wt{H}\ov{e}Q^3L^2}{acbedf}}  \,,
    \end{split} \\[2ex]
    \begin{split}
        \O^{(3)}_{\und{\wt{H}\ov{e}Q^3L^2}{acbfed}} &= +\, \O^{(1)}_{\und{\wt{H}\ov{e}Q^3L^2}{abedcf}} - \O^{(1)}_{\und{\wt{H}\ov{e}Q^3L^2}{abefcd}} - \O^{(1)}_{\und{\wt{H}\ov{e}Q^3L^2}{adebcf}} + \O^{(1)}_{\und{\wt{H}\ov{e}Q^3L^2}{adefcb}} + \O^{(1)}_{\und{\wt{H}\ov{e}Q^3L^2}{afebcd}} - \O^{(1)}_{\und{\wt{H}\ov{e}Q^3L^2}{afedcb}} \\
        &\phantom{=}\ + \O^{(2)}_{\und{\wt{H}\ov{e}Q^3L^2}{abcdef}} - \O^{(2)}_{\und{\wt{H}\ov{e}Q^3L^2}{abcfed}} - \O^{(2)}_{\und{\wt{H}\ov{e}Q^3L^2}{adcbef}} + \O^{(2)}_{\und{\wt{H}\ov{e}Q^3L^2}{adcfeb}} + \O^{(2)}_{\und{\wt{H}\ov{e}Q^3L^2}{afcbed}} - \O^{(2)}_{\und{\wt{H}\ov{e}Q^3L^2}{afcdeb}} \\
        &\phantom{=}\ - \O^{(3)}_{\und{\wt{H}\ov{e}Q^3L^2}{abcdfe}} + \O^{(3)}_{\und{\wt{H}\ov{e}Q^3L^2}{abcefd}} + \O^{(3)}_{\und{\wt{H}\ov{e}Q^3L^2}{acbdef}} \,.
    \end{split}
\end{align}
These relations show that every operator in $\fB'_{\wt{H}\ov{e}Q^3L^2}\setminus \fB''_{\wt{H}\ov{e}Q^3L^2}$ lies in the span of $\fB''_{\wt{H}\ov{e}Q^3L^2}$. Hence $\fB''_{\wt{H}\ov{e}Q^3L^2}$ spans the same space as the full orbit set $\fB'_{\wt{H}\ov{e}Q^3L^2}$. Since $\fB''_{\wt{H}\ov{e}Q^3L^2}$ is linearly independent and contains exactly $\dim W_{\wt{H}\ov{e}Q^3L^2}=25$ operators, it follows that
\begin{equation}
    \spa\!\left(\fB''_{\wt{H}\ov{e}Q^3L^2}\right)
    =
    \spa\!\left(\fB'_{\wt{H}\ov{e}Q^3L^2}\right)
    =
    W_{\wt{H}\ov{e}Q^3L^2} \,.
\end{equation}
Therefore, $\fB''_{\wt{H}\ov{e}Q^3L^2}$ is a basis of $W_{\wt{H}\ov{e}Q^3L^2}$, so the three-term set $\B_{\wt{H}\ov{e}Q^3L^2}$ is a basis for the operator type $\wt{H}\ov{e}Q^3L^2$. Because the number of terms agrees with the independent minimal count obtained from \texttt{Sym2Int}, this basis is minimal. This example illustrates the general strategy of our construction: the group-theoretic counting fixes the target dimension of the flavor-permutation space, the permutation-theoretic counting fixes the number of terms required in a minimal basis, and the explicit linear relations verify that the chosen terms indeed realize those requirements.

\section{Summary of results and classification}
\label{sec:classification}

We now summarize the main results of sections~\ref{sec:dim-6-ops}--\ref{sec:dim-15-ops}, whose central output is a collection of explicit operator bases. These sections cover all non-derivative baryon-number-violating operator types from dimensions 6 through 11, together with the symmetry-protected non-derivative operator types from dimensions 12 through 15. Almost all of the bases we construct are minimal, and the few exceptions are collected in appendix~\ref{app:nonminimal}.

The number of operator types at each dimension is listed in table~\ref{tab:summary}, where complex-conjugate types are not counted separately. Throughout, we choose the representative with $\Delta B>0$, with the corresponding operators with $\Delta B<0$ obtained by complex conjugation.

\begin{table}[b]
    \centering
    \renewcommand{\arraystretch}{1}
    \begin{tabular}{c c c c c} 
        \hline
        \multirow{2}{*}{Dimension} & \# non-deriv. & \# non-deriv. & \# non-deriv. & \# non-deriv. \\
        & BNV operator types &  $\Delta B = 1$ & $\Delta B = 2$ & $\Delta B = 3$ \\
        \hline\hline
        6 & 4 & 4 & 0 & 0 \\
        7 & 4 & 4 & 0 & 0 \\
        8 & 7 & 7 & 0 & 0 \\
        9 & 26 & 23 & 3 & 0 \\
        10 & 54 & 54 & 0 & 0 \\
        11 & 60 & 53 & 7 & 0 \\
        12 & 178 & 164 & 14 & 0 \\
        13 & 274 & 225 & 49 & 0\\
        14 & 353 & 317 & 36 & 0 \\
        15 & 791 & 614 & 168 & 9\\
        \hline
    \end{tabular}
    \caption{Numbers of non-derivative baryon-number-violating operator types through mass dimension 15. Complex-conjugate types are counted once.}
    \label{tab:summary}
\end{table}

Because we restrict attention to non-derivative baryon-number-violating operators, the classification is relatively straightforward. We organize the operators first by $(\Delta B,\Delta L)$, and then by schematic field content, which we refer to as the field-content class. When a given class contains many operator types, it is sometimes useful to refine it further into field-content subclasses. For example, $\ov{L}\ov{e}\psi^4\phi^2$ is a field-content subclass of the broader class $\ov{\ell}\ov{r}\psi^4\phi^2$. The resulting classification is shown in table~\ref{tab:classification}.

\begin{table}[p]
    \centering
    \renewcommand{\arraystretch}{1.3}
    \begin{tabular}{c c c c c} 
        \hline
        Dimension & $(\Delta B,\Delta L)$ & \multicolumn{2}{c}{Field-content class} & \# types \\
        \hline\hline
        6 & $(1,1)$ & $\psi^4$ & & 4 \\
        \hline
        7 & $(1,-1)$ & $\ov{\psi}\psi^3\phi$ & & 4 \\
        \hline
        8 & $(1,1)$ & $\psi^4\phi^2$ & & 7 \\
        \hline
        \multirow{4}{*}{9} & \multirow{2}{*}{$(1,-1)$} & $\ov{\psi}^2\psi^4$ & & 15 \\
         & & $\ov{\psi}\psi^3\phi^3$ & & 6 \\\cline{2-5}
         & $(1,3)$ & $\psi^6$ & & 2  \\\cline{2-5}
         & $(2,0)$ & $\psi^6$ & & 3  \\
        \hline 
        \multirow{7}{*}{10} & $(1,-3)$ & $\ov{\psi}^3\psi^3\phi$ & & 1 \\\cline{2-5}
         & \multirow{6}{*}{$(1,1)$} & \multirow{5}{*}{$\ov{\psi}\psi^5\phi$} & $\ov{L}\psi^5\phi$ & 10 \\
         & & & $\ov{e}\psi^5\phi$ & 10 \\
         & & & $\ov{Q}\psi^5\phi$ & 9 \\
         & & & $\ov{d}\psi^5\phi$ & 9 \\
         & & & $\ov{u}\psi^5\phi$ & 8 \\\cline{3-5}
         & & $\psi^4\phi^4$ & & 7 \\
        \hline
        \multirow{10}{*}{11} & \multirow{8}{*}{$(1,-1)$} & \multirow{2}{*}{$\ov{\ell}^2\psi^4\phi^2$} & $\ov{L}^2\psi^4\phi^2$ & 8 \\
         & & & $\ov{L}\ov{Q}\psi^4\phi^2$ & 6 \\\cline{3-5}
         & & \multirow{4}{*}{$\ov{\ell}\ov{r}\psi^4\phi^2$} & $\ov{L}\ov{e}\psi^4\phi^2$ & 6 \\
         & & & $\ov{L}\ov{d}\psi^4\phi^2$ & 3 \\
         & & & $\ov{L}\ov{u}\psi^4\phi^2$ & 5 \\
         & & & $\ov{Q}\ov{e}\psi^4\phi^2$ & 3 \\\cline{3-5}
         & & $\ov{r}^2\psi^4\phi^2$ & & 9 \\\cline{3-5}
         & & $\ov{\psi}\psi^3\phi^5$ & & 6 \\\cline{2-5}
         & $(1,3)$ & $\psi^6\phi^2$ & & 7 \\\cline{2-5}
         & $(2,0)$ & $\psi^6\phi^2$ & & 7 \\
        \hline
        \multirow{2}{*}{12} & $(2,-2)$ & $\ov{\psi}^2\psi^6$ & & 4 \\\cline{2-5}
         & $(2,2)$ & $\psi^8$ & & 10 \\
        \hline
        14 & $(1,5)$ & $\psi^8\phi^2$ & & 2 \\
        \hline
        \multirow{2}{*}{15} & $(1,-5)$ & $\ov{\psi}^5\psi^3\phi^3$ & & 1 \\\cline{2-5}
         & $(2,-4)$ & $\ov{\psi}^4\psi^6$ & & 1 \\\cline{2-5}
         & $(2,4)$ & $\psi^{10}$ & & 9 \\\cline{2-5}
         & $(3,1)$ & $\psi^{10}$ & & 9 \\\cline{2-5}
        \hline
    \end{tabular}
    \caption{Classification of the non-derivative baryon-number-violating operator types with explicit bases in sections~\ref{sec:dim-6-ops}--\ref{sec:dim-15-ops}.}
    \label{tab:classification}
\end{table}

\clearpage

\section{Comparison with the literature}
\label{sec:comparison}

Finally, before presenting our results in sections~\ref{sec:dim-6-ops}--\ref{sec:dim-15-ops}, we compare them with the existing literature, focusing in particular on works that claim to provide minimal bases for baryon-number-violating SMEFT operators. The goal of this section is to first assess the correctness of these proposed bases and determine whether they are in fact minimal in the sense defined in subsection~\ref{subsec:terminology}, then to clarify how our constructions relate to them. In many cases, we reproduce earlier results. In others, we use different but equivalent contractions in order to obtain bases that are simpler to present or more convenient to use. We also identify cases in which previously proposed bases are not in fact minimal, and show explicitly how our smaller bases span the corresponding operator spaces. In this way, the comparison both situates our results within the literature and highlights the improvements achieved by our construction.

\subsection{Dimension-6 and 7 operators}

Dimension-6 and dimension-7 operators have been studied extensively, and valid minimal bases at these dimensions were given in refs.~\cite{Grzadkowski:2010es} and \cite{Lehman:2014jma}, respectively. Our bases for dimension-6 (section~\ref{sec:dim-6-ops}) and dimension-7 (section~\ref{sec:dim-7-ops}) operators are taken directly from these works.

\subsection{Dimension-8 operators}

Comparing our bases for dimension-8 operators (section~\ref{sec:dim-8-ops}) to those of Murphy in ref.~\cite{Murphy:2020rsh}, we find agreement up to an overall sign for every operator type whose basis contains only a single term. For the operator type $\wt{H} u d Q L H$, Murphy chooses a basis
\begin{align}
    \O^{(M1)}_{\wt{H}udQLH} &= \eps_{\alpha\beta\gamma}\eps_{ij}(d^\alpha_aCu^\beta_b)(Q^{i\gamma}_cCL^j_d)(H^\dag H) \,, \\
    \O^{(M2)}_{\wt{H}udQLH} &= \eps_{\alpha\beta\gamma}(\eps\tau^I)_{ij}(d^\alpha_aCu^\beta_b)(Q^{i\gamma}_cCL^j_d)(H^\dag\tau^I H) \,.
\end{align}
Our first basis term, 
\begin{equation}
    \O^{(1)}_{\wt{H}udQLH} = \O^{(M1)}_{\wt{H}udQLH} \,,
\end{equation} 
coincides with Murphy's first term. However, we write our second basis term as
\begin{equation}
    \O^{(2)}_{\wt{H}udQLH} = \eps_{\alpha\beta\gamma}\eps_{ik}\eps_{jl}(d^\alpha_aCu^\beta_b)(Q^{i\gamma}_cCL^j_d)\wt{H}^kH^l
\end{equation}
since throughout this paper, we favor simpler contractions whenever possible. Murphy's second term $\O^{(M2)}_{\wt{H}udQLH}$ can then be expressed in our basis as
\begin{equation}
    \O^{(M2)}_{\wt{H}udQLH} = \O^{(1)}_{\wt{H}udQLH} - 2\O^{(2)}_{\wt{H}udQLH} \,,
\end{equation}
which shows the equivalence of the two bases.

For the operator type $\wt{H}Q^3LH$, Murphy proposes the basis
\begin{align}
    \O^{(M1)}_{\wt{H}Q^3LH} &= \eps_{\alpha\beta\gamma}\eps_{il}\eps_{jk}(Q^{i\alpha}_aCQ^{j\beta}_b)(Q^{k\gamma}_cCL^l_d)(H^\dag H) \,, \\
    \O^{(M2)}_{\wt{H}Q^3LH} &= \eps_{\alpha\beta\gamma}(\eps\tau^I)_{il}\eps_{jk}(Q^{i\alpha}_aCQ^{j\beta}_b)(Q^{k\gamma}_cCL^l_d)(H^\dag\tau^I H) \,, \\
    \O^{(M3)}_{\wt{H}Q^3LH} &= \eps_{\alpha\beta\gamma}\eps_{il}(\eps\tau^I)_{jk}(Q^{i\alpha}_aCQ^{j\beta}_b)(Q^{k\gamma}_cCL^l_d)(H^\dag\tau^I H) \,.
\end{align}
However, we find that this basis is not minimal and that only two terms are needed, as already noted in ref.~\cite{Fonseca:Mainz2023}. Our minimal basis is
\begin{align}
    \O^{(1)}_{\wt{H}Q^3LH} &= \eps_{\alpha\beta\gamma}\eps_{im}\eps_{jl}\eps_{kn}(Q^{i\alpha}_aCQ^{j\beta}_b)(Q^{k\gamma}_cCL^l_d)\wt{H}^m H^n \,, \\
    \O^{(2)}_{\wt{H}Q^3LH} &= \eps_{\alpha\beta\gamma}\eps_{ik}\eps_{jm}\eps_{ln}(Q^{i\alpha}_aCQ^{j\beta}_b)(Q^{k\gamma}_cCL^l_d)\wt{H}^m H^n \,.
\end{align}
To show that Murphy's operators lie in the span of our basis, we express them in terms of our operators as
\begin{align}
    \O^{(M1)}_{\und{\wt{H}Q^3 LH}{abcd}} &= - \O^{(1)}_{\und{\wt{H}Q^3 LH}{bacd}} + \O^{(1)}_{\und{\wt{H}Q^3 LH}{cabd}} - \O^{(1)}_{\und{\wt{H}Q^3 LH}{cbad}} - \O^{(2)}_{\und{\wt{H}Q^3 LH}{abcd}} + \O^{(2)}_{\und{\wt{H}Q^3 LH}{acbd}} - \O^{(2)}_{\und{\wt{H}Q^3 LH}{bacd}} \,, \\
    \O^{(M2)}_{\und{\wt{H}Q^3 LH}{abcd}} &= - \O^{(1)}_{\und{\wt{H}Q^3 LH}{bacd}} + \O^{(1)}_{\und{\wt{H}Q^3 LH}{cabd}} - \O^{(1)}_{\und{\wt{H}Q^3 LH}{cbad}} - \O^{(2)}_{\und{\wt{H}Q^3 LH}{abcd}} + \O^{(2)}_{\und{\wt{H}Q^3 LH}{acbd}} + \O^{(2)}_{\und{\wt{H}Q^3 LH}{bacd}} \,, \\
    \O^{(M3)}_{\und{\wt{H}Q^3 LH}{abcd}} &= +\O^{(1)}_{\und{\wt{H}Q^3 LH}{bacd}} + \O^{(1)}_{\und{\wt{H}Q^3 LH}{cabd}} - \O^{(1)}_{\und{\wt{H}Q^3 LH}{cbad}} - \O^{(2)}_{\und{\wt{H}Q^3 LH}{abcd}} + \O^{(2)}_{\und{\wt{H}Q^3 LH}{acbd}} - \O^{(2)}_{\und{\wt{H}Q^3 LH}{bacd}} \,.
\end{align}
Therefore, two terms are sufficient to span the operator space.

\subsection{Dimension-9 operators}

Liao and Ma~\cite{Liao:2020jmn} provide a minimal basis for all dimension-9 operators. We verify that all of their bases for non-derivative baryon-number-violating operators are indeed minimal, and we adopt their contractions for most operator types. However, in keeping with our goal of presenting the simplest possible bases, for some operator types we instead use different but equivalent bases. One example is the operator type $\mathcal{T}=\ov{L}\ov{Q}ud^3$, for which we choose
\begin{align}
    \O_{\ov{L}\ov{Q}ud^3}^{(1)} &= \eps_{\alpha\gamma\sigma}\delta_{\beta\rho}\eps_{ij}(\ov{L}^i_ad^\alpha_b)(\ov{Q}^{j\beta}_cd^\gamma_d)(d^\rho_eCu^\sigma_f) \,, \\
    \O_{\ov{L}\ov{Q}ud^3}^{(2)} &= \eps_{\alpha\gamma\sigma}\delta_{\beta\rho}\eps_{ij}(\ov{L}^i_ad^\alpha_b)(\ov{Q}^{j\beta}_cu^\gamma_d)(d^\rho_eCd^\sigma_f) \,, \\
    \O_{\ov{L}\ov{Q}ud^3}^{(3)} &= \eps_{\alpha\gamma\sigma}\delta_{\beta\rho}\eps_{ij}(\ov{L}^i_au^\alpha_b)(\ov{Q}^{j\beta}_cd^\gamma_d)(d^\rho_eCd^\sigma_f) \,,
\end{align}
whereas the basis given in ref.~\cite{Liao:2020jmn} is
\begin{align}
    \O_{\ov{L}\ov{Q}ud^3}^{(L1)} &= \delta_{\alpha\beta}\eps_{\gamma\sigma\rho}\eps_{ij}(\ov{L}^i_ad^\alpha_b)(\ov{Q}^{j\beta}_cd^\gamma_d)(d^\rho_eCu^\sigma_f) \,, \\
    \O_{\ov{L}\ov{Q}ud^3}^{(L2)} &= \eps_{\alpha\rho\sigma}\delta_{\beta\gamma}\eps_{ij}(\ov{L}^i_ad^\alpha_b)(\ov{Q}^{j\beta}_cu^\gamma_d)(d^\rho_eCd^\sigma_f) \,, \\
    \O_{\ov{L}\ov{Q}ud^3}^{(L3)} &= \delta_{\alpha\rho}\eps_{\beta\gamma\sigma}\eps_{ij}(\ov{L}^i_a\sigma_{\mu\nu}Q^{\C,j\alpha}_b)(d^\beta_cC\sigma^{\mu\nu}d^\gamma_d)(d^\rho_eCu^\sigma_f) \,,
\end{align}
where $\sigma^{\mu\nu} = i\gamma^\mu\gamma^\nu - ig^{\mu\nu}$ is the antisymmetric Lorentz generator in the Dirac spinor representation. Although Liao and Ma's basis is valid and minimal, it includes the tensor structure $\O_{\ov{L}\ov{Q}ud^3}^{(L3)}$, whereas our basis is written entirely in terms of scalar bilinears. Our choice is therefore more compact and more convenient for explicit manipulations. The equivalence of the two bases may be exhibited explicitly through
\begin{align}
    \O_{\und{\ov{L}\ov{Q}ud^3}{abcdef}}^{(L1)} &= +\O_{\und{\ov{L}\ov{Q}ud^3}{aecdbf}}^{(1)} - \O_{\und{\ov{L}\ov{Q}ud^3}{abcedf}}^{(3)} \,, \\
    \O_{\und{\ov{L}\ov{Q}ud^3}{abcdef}}^{(L2)} &= -\O_{\und{\ov{L}\ov{Q}ud^3}{abcedf}}^{(1)} + \O_{\und{\ov{L}\ov{Q}ud^3}{abcdef}}^{(2)} \,, \\
    \O_{\und{\ov{L}\ov{Q}ud^3}{abcdef}}^{(L3)} &= + \O_{\und{\ov{L}\ov{Q}ud^3}{abcdef}}^{(1)} + \O_{\und{\ov{L}\ov{Q}ud^3}{adcbef}}^{(1)} \,.
\end{align}
Hence, the two sets span the same operator space, but our basis provides a simpler choice of representatives.

\begin{table}[t!]
    \centering
    \renewcommand{\arraystretch}{1.2}
    \begin{tabular}{lcc}
         Operator type & Our \#terms & He and Ma \#terms \\ [0.5ex] 
         \hline
         $u^4d^2e^2$ & 2 & 3 ($2+0+1$) \\
         $u^3d^3L^2$ & 2 & 2 ($2+0+0$) \\
         $u^3d^2QeL$ & 3 & 3 ($0+3+0$) \\
         $u^3dQ^2e^2$ & 2 & 3 ($1+0+2$) \\
         $u^2d^2Q^2L^2$ & $5^*$ & 6 ($3+0+3$) \\
         $u^2dQ^3eL$ & 4 & 3 ($0+3+0$) \\
         $u^2Q^4e^2$ & 2 & 1 ($1+0+0$) \\
         $udQ^4L^2$ & 4 & 5 ($2+0+3$) \\
         $uQ^5eL$ & 2 & 1 ($0+1+0$) \\
         $Q^6L^2$ & 2 & 2 ($1+0+1$) \\
    \end{tabular}
    \caption{Comparison of the number of terms in our bases and in those of He and Ma~\cite{He:2021mrt} for dimension-12 operators with $(\Delta B,\Delta L)=(2,2)$. For He and Ma, the numbers in parentheses give the decomposition into scalar-, vector-, and tensor-current operators. $^*$A minimal basis for $u^2d^2Q^2L^2$ only contains 4 terms; see appendix~\ref{app:nonminimal}.}
    \label{tab:dim-12-comparison}
\end{table}

\subsection{Dimension-10 to 15 operators}

For dimensions 10 through 15, we are not aware of earlier work devoted specifically to explicit minimal bases for fully flavored baryon-number-violating SMEFT operators in the sense used here. He and Ma~\cite{He:2021mrt} study the subset of dimension-12 operators relevant to dinucleon-to-dilepton decays and present a basis for that restricted problem. Since their analysis assumes first-generation quarks, while allowing unrestricted lepton flavor, a direct one-to-one comparison with our fully flavored bases is not possible. Nevertheless, their operator space is a restriction of the full space considered here, so one would generally expect the number of basis terms required in their setup not to exceed that required in ours. Table~\ref{tab:dim-12-comparison} shows that this expectation fails for several operator types, for which He and Ma use more terms than appear in our bases. This suggests that their chosen representatives are not always minimal. We also note that many operators in ref.~\cite{He:2021mrt} are written in terms of vector or tensor currents, whereas throughout this work we favor equivalent representatives built from scalar bilinears whenever possible.

\section{Dimension-6 operators}
\label{sec:dim-6-ops}

\begin{center}
\renewcommand{\arraystretch}{1.3}
\begin{longtable}{l l} 
 \hline
 Operator type & Terms \\ [0.5ex] 
 \hline\hline
 $u^2de$ & $\eps_{\alpha\beta\gamma}(d^\alpha_aCu^\beta_b)(u^\gamma_cCe_d)$ \\
 $udQL$ & $\eps_{\alpha\beta\gamma}\eps_{ij}(d^\alpha_aCu^\beta_b)(Q^{i\gamma}_cCL^j_d)$ \\
 $uQ^2e$ & $\eps_{\alpha\beta\gamma}\eps_{ij}(Q^{i\alpha}_aCQ^{j\beta}_b)(u^\gamma_c Ce_d)$ \\
 $Q^3L$ & $\eps_{\alpha\beta\gamma}\eps_{il}\eps_{jk}(Q^{i\alpha}_aCQ^{j\beta}_b)(Q^{k\gamma}_cCL^l_d)$ \\
 \hline
\end{longtable}
\end{center}

\section{Dimension-7 operators}
\label{sec:dim-7-ops}

\begin{center}
\renewcommand{\arraystretch}{1.3}
\begin{longtable}{l l} 
 \hline
 Operator type & Terms \\ [0.5ex] 
 \hline\hline
 $\wt{H}\ov{L}ud^2$ & $\eps_{\alpha\beta\gamma}(\ov{L}_ad^\alpha_b\wt{H})(u^\beta_cCd^\gamma_d)$ \\
 $\wt{H}\ov{L}dQ^2$ & $\eps_{\alpha\beta\gamma}\delta_{ij}\eps_{kl}(\ov{L}^i_ad^\alpha_b)(Q^{j\beta}_cCQ^{k\gamma}_d)\wt{H}^l$ \\
 $\wt{H}\ov{e}d^2Q$ & $\eps_{\alpha\beta\gamma}(H^\dag\ov{e}_aQ^{\alpha}_b)(d^\beta_cCd^\gamma_d)$ \\
 $\ov{L}d^3H$ & $\eps_{\alpha\beta\gamma}(\ov{L}_ad^\alpha_bH)(d^\beta_cCd^\gamma_d)$ \\
 \hline
\end{longtable}
\end{center}

\section{Dimension-8 operators}
\label{sec:dim-8-ops}

\begin{center}
\renewcommand{\arraystretch}{1.3}
\begin{longtable}{l l} 
 \hline
 Operator type & Terms \\ [0.5ex] 
 \hline\hline
 $\wt{H}^2u^2QL$ & $\eps_{\alpha\beta\gamma}\eps_{ik}\eps_{jl}(u^\alpha_aCu^\beta_b)(Q^{i\gamma}_cCL^j_d)\widetilde{H}^k\widetilde{H}^l$ \\
 $\wt{H}u^2deH$ & $\eps_{\alpha\beta\gamma}(d^\alpha_aCu^\beta_b)(u^\gamma_cCe_d)(H^\dag H)$ \\
 \hline
 \multirow{2}{*}{$\wt{H}udQLH$} & $\eps_{\alpha\beta\gamma}\eps_{ij}(d^\alpha_aCu^\beta_b)(Q^{i\gamma}_cCL^j_d)(H^\dag H)$ \\
  & $\eps_{\alpha\beta\gamma}\eps_{ik}\eps_{jl}(d^\alpha_aCu^\beta_b)(Q^{i\gamma}_cCL^j_d)\wt{H}^kH^l$ \\
 \hline
 $\wt{H}uQ^2eH$ & $\eps_{\alpha\beta\gamma}\eps_{ik}\eps_{jl}(e_aCu^\alpha_b)(Q^{i\beta}_cCQ^{j\gamma}_d)\wt{H}^k H^l$ \\
 \hline
 \multirow{2}{*}{$\wt{H}Q^3LH$} & $\eps_{\alpha\beta\gamma}\eps_{im}\eps_{jl}\eps_{kn}(Q^{i\alpha}_aCQ^{j\beta}_b)(Q^{k\gamma}_cCL^l_d)\wt{H}^m H^n$ \\
  & $\eps_{\alpha\beta\gamma}\eps_{ik}\eps_{jm}\eps_{ln}(Q^{i\alpha}_aCQ^{j\beta}_b)(Q^{k\gamma}_cCL^l_d)\wt{H}^m H^n$ \\
 \hline
 $d^2QLH^2$ & $\eps_{\alpha\beta\gamma}\eps_{ik}\eps_{jl}(d^\alpha_aCd^\beta_b)(Q^{i\gamma}_cCL^j_d)H^kH^l$ \\
 $dQ^2eH^2$ & $\eps_{\alpha\beta\gamma}\eps_{ik}\eps_{jl}(e_aCd^\alpha_b)(Q^{i\beta}_cCQ^{j\gamma}_d)H^kH^l$ \\
 \hline
\end{longtable}
\end{center}

\vspace{-5ex}
\section{Dimension-9 operators}
\label{sec:dim-9-ops}

\subsection{\texorpdfstring{$(\Delta B, \Delta L) = (1, -1), \ov{\psi}^2\psi^4$}{1, -1, psi6}}
\label{subsec:1, -1, psi6}

\begin{center}
\renewcommand{\arraystretch}{1.3}
\begin{longtable}{l l} 
 \hline
 Operator type & Terms \\ [0.5ex] 
 \hline\hline
 \multirow{2}{*}{$\ov{L}^2ud^2e$} & $\eps_{\alpha\beta\gamma}\eps_{ij}(\ov{L}^i_ae_b)(\ov{L}^j_cd^\alpha_d)(d^\beta_eCu^\gamma_f)$ \\
  & $\eps_{\alpha\beta\gamma}\eps_{ij}(e_aCd^\alpha_b)(\ov{L}^i_cd^\beta_d)(\ov{L}^j_eu^\gamma_f)$ \\
 \hline
 $\ov{L}^2d^2QL$ & $\eps_{\alpha\beta\gamma}\delta_{ik}\delta_{jl}(L^i_aCQ^{j\alpha}_b)(\ov{L}^k_cd^{\beta}_d)(\ov{L}^l_ed^\gamma_f)$ \\
 $\ov{L}^2dQ^2e$ & $\eps_{\alpha\beta\gamma}\delta_{ik}\delta_{jl}(\ov{L}^i_ae_b)(\ov{L}^j_cd^\alpha_d)(Q^{k\beta}_eCQ^{l\gamma}_f)$ \\
 $\ov{L}\ov{e}d^3L$ & $\eps_{\alpha\beta\gamma}\delta_{ij}(\ov{e}_aL^i_b)(\ov{L}^j_cd^\alpha_d)(d^\beta_eCd^\gamma_f)$ \\
 $\ov{L}\ov{e}d^2Qe$ & $\eps_{\alpha\beta\gamma}\delta_{ij}(\ov{e}_aQ^{i\alpha}_b)(\ov{L}^j_cd^\beta_d)(e_eCd^\gamma_f)$ \\
 \hline
 \multirow{3}{*}{$\ov{L}\ov{Q}ud^3$} & $\eps_{\alpha\gamma\sigma}\delta_{\beta\rho}\eps_{ij}(\ov{L}^i_ad^\alpha_b)(\ov{Q}^{j\beta}_cd^\gamma_d)(d^\rho_eCu^\sigma_f)$ \\
\hline
\end{longtable}
\end{center}

\subsection{\texorpdfstring{$(\Delta B,\Delta L) = (1,-1), \ov{\psi}\psi^3\varphi^3$}{1, -1, psi4H3}}
\label{subsec:1, -1, psi4H3}

\begin{center}
\renewcommand{\arraystretch}{1.3}
\begin{longtable}{l l} 
 \hline
 Operator type & Terms \\ [0.5ex] 
 \hline\hline
 $\wt{H}^2\ov{L}ud^2H$ & $\eps_{\alpha\beta\gamma}(\ov{L}_ad^\alpha_b\wt{H})(d^\beta_cCu^\gamma_d)(H^\dag H)$ \\
 \hline
 \multirow{2}{*}{$\wt{H}^2\ov{L}dQ^2H$} & $\eps_{\alpha\beta\gamma}\delta_{ij}\eps_{kl}(\ov{L}^i_ad^\alpha_b)(Q^{j\beta}_cCQ^{k\gamma}_d)\wt{H}^l(H^\dag H)$ \\
  & $\eps_{\alpha\beta\gamma}\eps_{ik}\eps_{jl}(\ov{L}_ad^\alpha_bH)(Q^{i\beta}_cCQ^{j\gamma}_d)\wt{H}^k\wt{H}^l$ \\
 \hline
 $\wt{H}^2\ov{e}d^2QH$ & $\eps_{\alpha\beta\gamma}(H^\dag\ov{e}_aQ^\alpha_b)(d^\beta_cCd^\gamma_d)(H^\dag H)$ \\
 $\wt{H}\ov{L}d^3H^2$ & $\eps_{\alpha\beta\gamma}(\ov{L}_ad^\alpha_bH)(d^\beta_cCd^\gamma_d)(H^\dag H)$ \\
 $\wt{H}^3\ov{e}Q^3$ & $\eps_{\alpha\beta\gamma}\eps_{ik}\eps_{jl}(H^\dag\ov{e}_aQ^{\alpha}_b)(Q^{i\beta}_cCQ^{j\gamma}_d)\wt{H}^k\wt{H}^l$ \\
 $\wt{H}^3\ov{L}uQ^2$ & $\eps_{\alpha\beta\gamma}\eps_{ik}\eps_{jl}(\ov{L}_au^\alpha_b\wt{H})(Q^{i\beta}_cCQ^{j\gamma}_d)\wt{H}^k\wt{H}^l$ \\
 \hline
\end{longtable}
\end{center}

\subsection{\texorpdfstring{$(\Delta B, \Delta L) = (1, 3), \psi^6$}{1, 3, psi6}}
\label{subsec:1, 3, psi6}

\begin{center}
\renewcommand{\arraystretch}{1.3}
\begin{longtable}{l l} 
 \hline
 Operator type & Terms \\ [0.5ex] 
 \hline\hline
$u^3eL^2$ & $\eps_{\alpha\beta\gamma}\eps_{ij}(e_aCu^\alpha_b)(u^\beta_cCu^\gamma_d)(L^i_eCL^j_f)$ \\
 $u^2QL^3$ & $\eps_{\alpha\beta\gamma}\eps_{ik}\eps_{jl}(u^\alpha_aCu^\beta_b)(Q^{i\gamma}_cCL^j_d)(L^k_eCL^l_f)$ \\
 \hline
\end{longtable}
\end{center}

\subsection{\texorpdfstring{$(\Delta B, \Delta L) = (2,0), \psi^6$}{2, 0, psi6}}
\label{subsec:2, 0, psi6}

\begin{center}
\renewcommand{\arraystretch}{1.3}
\begin{longtable}{l l} 
 \hline
 Operator type & Terms \\ [0.5ex] 
 \hline\hline
 \multirow{2}{*}{$u^2d^4$} & $\eps_{\alpha\gamma\sigma}\eps_{\beta\rho\tau}(d^\alpha_aCu^\beta_b)(d^\gamma_cCu^\rho_d)(d^\sigma_eCd^\tau_f)$ \\
  & $\eps_{\alpha\rho\sigma}\eps_{\beta\gamma\tau}(d^\alpha_aCu^\beta_b)(d^\gamma_cCu^\rho_d)(d^\sigma_eCd^\tau_f)$ \\
 \hline
 $ud^3Q^2$ & $\eps_{\alpha\gamma\sigma}\eps_{\beta\rho\tau}\eps_{ij}(u^\alpha_aCd^\beta_b)(d^\gamma_cCd^\rho_d)(Q^{i\sigma}_eCQ^{j\tau}_f)$ \\
 \hline
 \multirow{2}{*}{$d^2Q^4$} & $\eps_{\alpha\gamma\sigma}\eps_{\beta\rho\tau}\eps_{ik}\eps_{jl}(d^\alpha_aCd^\beta_b)(Q^{i\gamma}_cCQ^{j\rho}_d)(Q^{k\sigma}_eCQ^{l\tau}_f)$ \\
  & $\eps_{\alpha\rho\sigma}\eps_{\beta\gamma\tau}\eps_{ik}\eps_{jl}(d^\alpha_aCd^\beta_b)(Q^{i\gamma}_cCQ^{j\rho}_d)(Q^{k\sigma}_eCQ^{l\tau}_f)$ \\
 \hline
\end{longtable}
\end{center}

\clearpage

\section{Dimension-10 operators}
\label{sec:dim-10-ops}

\subsection{\texorpdfstring{$(\Delta B, \Delta L) = (1, -3), \ov{\psi}^3\psi^3\varphi$}{1, -3, psi6H}}
\label{subsec:1, -3, psi6H}

\begin{center}
\renewcommand{\arraystretch}{1.3}
\begin{longtable}{l l} 
 \hline
 Operator type & Terms \\ [0.5ex] 
 \hline\hline
 $\wt{H}\ov{L}^3d^3$ & $\eps_{\alpha\beta\gamma}\eps_{ij}(\ov{L}_ad^\alpha_b\wt{H})(\ov{L}^i_cd^\beta_d)(\ov{L}^j_ed^\gamma_f)$ \\
 \hline
\end{longtable}
\end{center}

\subsection{\texorpdfstring{$(\Delta B,\Delta L) = (1,1), \ov{\psi}\psi^5\varphi$}{psi6H}}
\label{subsec:psi6H}

\subsubsection{\texorpdfstring{$\ov{L}\psi^5\varphi$}{L*psi5H}}
\label{subsubsec:L*psi5H}

\begin{center}
\renewcommand{\arraystretch}{1.3}
\begin{longtable}{l l} 
 \hline
 Operator type & Terms \\ [0.5ex] 
 \hline\hline
 $\wt{H}\ov{L}u^3e^2$ & $\eps_{\alpha\beta\gamma}(\ov{L}_au^\alpha_b\wt{H})(e_cCu^\beta_d)(e_eCu^\gamma_f)$ \\
 $\wt{H}\ov{L}u^2dL^2$ & $\eps_{\alpha\beta\gamma}\delta_{ij}\eps_{kl}(\ov{L}^i_au^\alpha_b)(u^\beta_cCd^\gamma_d)(L^j_eCL^k_f)\wt{H}^l$ \\
 \hline
 \multirow{2}{*}{$\wt{H}\ov{L}u^2QeL$} & $\eps_{\alpha\beta\gamma}\delta_{ij}\eps_{kl}(\ov{L}^i_au^\alpha_b)(e_cCu^\beta_d)(L^j_eCQ^{k\gamma}_f)\wt{H}^l$ \\
  & $\eps_{\alpha\beta\gamma}\delta_{ik}\eps_{jl}(\ov{L}^i_au^\alpha_b)(e_cCu^\beta_d)(L^j_eCQ^{k\gamma}_f)\wt{H}^l$ \\
 \hline
 \multirow{3}{*}{$\wt{H}\ov{L}uQ^2L^2$} & $\eps_{\alpha\beta\gamma}\delta_{ij}\eps_{kl}\eps_{mn}(\ov{L}^i_au^\alpha_b)(L^j_cCQ^{k\beta}_d)(L^l_eCQ^{m\gamma}_f)\wt{H}^n$ \\
  & $\eps_{\alpha\beta\gamma}\delta_{ik}\eps_{jn}\eps_{lm}(\ov{L}^i_au^\alpha_b)(L^j_cCQ^{k\beta}_d)(L^l_eCQ^{m\gamma}_f)\wt{H}^n$ \\
  & $\eps_{\alpha\beta\gamma}\delta_{in}\eps_{jl}\eps_{km}(\ov{L}^i_au^\alpha_b)(L^j_cCQ^{k\beta}_d)(L^l_eCQ^{m\gamma}_f)\wt{H}^n$ \\
\hline
\end{longtable}
\end{center}

\subsubsection{\texorpdfstring{$\ov{e}\psi^5\varphi$}{e*psi5H}}
\label{subsubsec:e*psi5H}

\begin{center} 
\renewcommand{\arraystretch}{1.3}
\begin{longtable}{l l} 
 \hline
 Operator type & Terms \\ [0.5ex] 
 \hline\hline
 $\wt{H}\ov{e}u^2deL$ & $\eps_{\alpha\beta\gamma}(H^\dag\ov{e}_aL_b)(e_cCu^\alpha_d)(u^\beta_eCd^\gamma_f)$ \\
 $\wt{H}\ov{e}u^2Qe^2$ & $\eps_{\alpha\beta\gamma}(H^\dag\ov{e}_aQ^{\alpha}_b)(e_cCu^\beta_d)(e_eCu^\gamma_f)$ \\
 \hline
 \multirow{2}{*}{$\wt{H}\ov{e}udQL^2$} & $\eps_{\alpha\beta\gamma}\eps_{ij}(H^\dag\ov{e}_aL_b)(L^i_cCQ^{j\alpha}_d)(d^\beta_eCu^\gamma_f)$ \\
  & $\eps_{\alpha\beta\gamma}\eps_{ij}\eps_{kl}(\ov{e}_aL^i_b)(L^j_cCQ^{k\alpha}_d)(d^\beta_eCu^\gamma_f)\wt{H}^l$ \\
 \hline
 \multirow{2}{*}{$\wt{H}\ov{e}uQ^2eL$} & $\eps_{\alpha\beta\gamma}\eps_{ij}(H^\dag\ov{e}_aQ^\alpha_b)(L^i_cCQ^{j\beta}_d)(e_eCu^\gamma_f)$ \\
  & $\eps_{\alpha\beta\gamma}\eps_{ij}\eps_{kl}(\ov{e}_aQ^{i\alpha}_b)(L^j_cCQ^{k\beta}_d)(e_eCu^\gamma_f)\wt{H}^l$ \\
 \hline
 \multirow{3}{*}{$\wt{H}\ov{e}Q^3L^2$} & $\eps_{\alpha\beta\gamma}\eps_{ij}\eps_{kl}\eps_{mn}(\ov{e}_aQ^{i\alpha}_b)(L^j_cCQ^{k\beta}_d)(L^l_eCQ^{m\gamma}_f)\wt{H}^n$ \\
\hline
\end{longtable}
\end{center}

\vspace{-4ex}

\subsubsection{\texorpdfstring{$\ov{Q}\psi^5\varphi$}{Q*psi5H}}
\label{subsubsec:Q*psi5H}

\begin{center} 
\renewcommand{\arraystretch}{1.3}
\begin{longtable}{l l} 
 \hline
 Operator type & Terms \\ [0.5ex] 
 \hline\hline
 \multirow{3}{*}{$\wt{H}\ov{Q}u^3de$} & $\delta_{\alpha\sigma}\eps_{\beta\gamma\rho}(\ov{Q}^{\alpha}_au^\beta_b\wt{H})(e_cCu^\gamma_d)(u^\rho_eCd^\sigma_f)$ \\
  & $\delta_{\alpha\gamma}\eps_{\beta\rho\sigma}(\ov{Q}^{\alpha}_au^\beta_b\wt{H})(e_cCu^\gamma_d)(u^\rho_eCd^\sigma_f)$ \\
  & $\delta_{\alpha\beta}\eps_{\gamma\rho\sigma}(\ov{Q}^{\alpha}_au^\beta_b\wt{H})(e_cCu^\gamma_d)(u^\rho_eCd^\sigma_f)$ \\
 \hline
 \multirow{6}{*}{$\wt{H}\ov{Q}u^2dQL$} & $\delta_{\alpha\sigma}\eps_{\beta\gamma\rho}\eps_{ij}(\ov{Q}^{\alpha}_au^\beta_b\wt{H})(u^\gamma_cCd^\rho_d)(Q^{i\sigma}_eCL^j_f)$ \\
  & $\delta_{\alpha\rho}\eps_{\beta\gamma\sigma}\eps_{ij}(\ov{Q}^{\alpha}_au^\beta_b\wt{H})(u^\gamma_cCd^\rho_d)(Q^{i\sigma}_eCL^j_f)$ \\
  & $\delta_{\alpha\gamma}\eps_{\beta\rho\sigma}\eps_{ij}(\ov{Q}^{\alpha}_au^\beta_b\wt{H})(u^\gamma_cCd^\rho_d)(Q^{i\sigma}_eCL^j_f)$ \\
  & $\delta_{\alpha\sigma}\eps_{\beta\gamma\rho}\delta_{ik}\eps_{jl}(\ov{Q}^{i\alpha}_au^\beta_b)(u^\gamma_cCd^\rho_d)(Q^{j\sigma}_eCL^k_f)\wt{H}^l$ \\
\hline
\end{longtable}
\end{center}

\subsubsection{\texorpdfstring{$\ov{d}\psi^5\varphi$}{d*psi5H}}
\label{subsubsec:d*psi5H}

\begin{center} 
\renewcommand{\arraystretch}{1.3}
\begin{longtable}{l l} 
 \hline
 Operator type & Terms \\ [0.5ex] 
 \hline\hline
 \multirow{2}{*}{$\wt{H}\ov{d}u^2d^2L$} & $\delta_{\alpha\sigma}\eps_{\beta\gamma\rho}(H^\dag\ov{d}^\alpha_aL_b)(d^\beta_cCu^\gamma_d)(d^\rho_eCu^\sigma_f)$ \\
  & $\delta_{\alpha\rho}\eps_{\beta\gamma\sigma}(H^\dag\ov{d}^\alpha_aL_b)(d^\beta_cCu^\gamma_d)(d^\rho_eCu^\sigma_f)$ \\
 \hline
 \multirow{3}{*}{$\wt{H}\ov{d}u^2dQe$} & $\delta_{\alpha\sigma}\eps_{\beta\gamma\rho}(H^\dag\ov{d}^\alpha_aQ^\beta_b)(d^\gamma_cCu^\rho_d)(u^\sigma_eCe_f)$ \\
  & $\delta_{\alpha\rho}\eps_{\beta\gamma\sigma}(H^\dag\ov{d}^\alpha_aQ^\beta_b)(d^\gamma_cCu^\rho_d)(u^\sigma_eCe_f)$ \\
  & $\delta_{\alpha\gamma}\eps_{\beta\rho\sigma}(H^\dag\ov{d}^\alpha_aQ^\beta_b)(d^\gamma_cCu^\rho_d)(u^\sigma_eCe_f)$ \\
 \hline
 \multirow{6}{*}{$\wt{H}\ov{d}udQ^2L$} & $\delta_{\alpha\sigma}\eps_{\beta\gamma\rho}\eps_{ij}(H^\dag\ov{d}^\alpha_aQ^\beta_b)(d^\gamma_cCu^\rho_d)(Q^{i\sigma}_eCL^j_f)$ \\
  & $\delta_{\alpha\rho}\eps_{\beta\gamma\sigma}\eps_{ij}(H^\dag\ov{d}^\alpha_aQ^\beta_b)(d^\gamma_cCu^\rho_d)(Q^{i\sigma}_eCL^j_f)$ \\
\hline
\end{longtable}
\end{center}

\newpage

\subsubsection{\texorpdfstring{$\ov{u}\psi^5\varphi$}{u*psi5H}}
\label{subsubsec:u*psi5H}

\begin{center}
\renewcommand{\arraystretch}{1.3}
\begin{longtable}{l l} 
 \hline
 Operator type & Terms \\ [0.5ex] 
 \hline\hline
 $\wt{H}\ov{u}u^3dL$ & $\delta_{\alpha\beta}\eps_{\gamma\rho\sigma}(H^\dag\ov{u}^\alpha_aL_b)(u^\beta_cCu^\gamma_d)(u^\rho_eCd^\sigma_f)$ \\
 $\wt{H}\ov{u}u^3Qe$ & $\delta_{\alpha\gamma}\eps_{\beta\rho\sigma}(H^\dag\ov{u}^\alpha_aQ^\beta_b)(u^\gamma_cCu^\rho_d)(u^\sigma_eCe_f)$ \\
 \hline
 \multirow{4}{*}{$\wt{H}\ov{u}u^2Q^2L$} & $\delta_{\alpha\rho}\eps_{\beta\gamma\sigma}\eps_{ij}(H^\dag\ov{u}^\alpha_a Q^\beta_b)(u^\gamma_cCu^\rho_d)(Q^{i\sigma}_eCL^j_f)$ \\
  & $\delta_{\alpha\beta}\eps_{\gamma\rho\sigma}\eps_{ij}(H^\dag\ov{u}^\alpha_a Q^\beta_b)(u^\gamma_cCu^\rho_d)(Q^{i\sigma}_eCL^j_f)$ \\
  & $\delta_{\alpha\rho}\eps_{\beta\gamma\sigma}\eps_{ij}\eps_{kl}(\ov{u}^\alpha_a Q^{i\beta}_b)(u^\gamma_cCu^\rho_d)(Q^{j\sigma}_eCL^k_f)\wt{H}^l$ \\
  & $\delta_{\alpha\beta}\eps_{\gamma\rho\sigma}\eps_{ij}\eps_{kl}(\ov{u}^\alpha_a Q^{i\beta}_b)(u^\gamma_cCu^\rho_d)(Q^{j\sigma}_eCL^k_f)\wt{H}^l$ \\
 \hline
 \multirow{2}{*}{$\ov{u}u^2d^2LH$} & $\delta_{\alpha\sigma}\eps_{\beta\gamma\rho}(\wt{H}^\dag\ov{u}^\alpha_aL_b)(d^\beta_cCu^\gamma_d)(d^\rho_eCu^\sigma_f)$ \\
\hline
\end{longtable}
\end{center}

\vspace{-5ex}

\subsection{\texorpdfstring{$(\Delta B,\Delta L) = (1,1), \psi^4\varphi^4$}{psi4H4}}
\label{subsec:psi4H4}

\begin{center}
\renewcommand{\arraystretch}{1.3}
\begin{longtable}{l l} 
 \hline
 Operator type & Terms \\ [0.5ex] 
 \hline\hline
 $\wt{H}^3u^2QLH$ & $\eps_{\alpha\beta\gamma}\eps_{ik}\eps_{jl}(L^i_aCQ^{j\alpha}_b)(u^\beta_cCu^\gamma_d)\wt{H}^k\wt{H}^l(H^\dag H)$ \\
 $\wt{H}^2u^2deH^2$ & $\eps_{\alpha\beta\gamma}(d^\alpha_aCu^\beta_b)(u^\gamma_cCe_d)(H^\dag H)(H^\dag H)$ \\
 \hline
 \multirow{2}{*}{$\wt{H}^2udQLH^2$} & $\eps_{\alpha\beta\gamma}\eps_{ij}(L^i_aCQ^{j\alpha}_b)(d^\beta_cCu^\gamma_d)(H^\dag H)(H^\dag H)$ \\
  & $\eps_{\alpha\beta\gamma}\eps_{ik}\eps_{jl}(L^i_aCQ^{j\alpha}_b)(d^\beta_cCu^\gamma_d)\wt{H}^kH^l(H^\dag H)$ \\
 \hline
 $\wt{H}^2uQ^2eH^2$ & $\eps_{\alpha\beta\gamma}\eps_{ik}\eps_{jl}(e_aCQ^{i\alpha}_b)(Q^{j\beta}_cCu^\gamma_d)\wt{H}^kH^l(H^\dag H)$ \\
 \hline
 \multirow{2}{*}{$\wt{H}^2Q^3LH^2$} & $\eps_{\alpha\beta\gamma}\eps_{im}\eps_{jp}\eps_{kn}\eps_{lq}(Q^{i\alpha}_aCQ^{j\beta}_b)(Q^{k\gamma}_cCL^l_d)\wt{H}^m\wt{H}^nH^pH^q$ \\
  & $\eps_{\alpha\beta\gamma}\eps_{il}\eps_{jm}\eps_{kn}(Q^{i\alpha}_aCQ^{j\beta}_b)(Q^{k\gamma}_cCL^l_d)\wt{H}^mH^n(H^\dag H)$ \\
 \hline
 $\wt{H}d^2QLH^3$ & $\eps_{\alpha\beta\gamma}\eps_{ik}\eps_{jl}(L^i_aCQ^{j\alpha}_b)(d^\beta_cCd^\gamma_d)H^kH^l(H^\dag H)$ \\
 $\wt{H}dQ^2eH^3$ & $\eps_{\alpha\beta\gamma}\eps_{ik}\eps_{jl}(e_aCd^\alpha_b)(Q^{i\beta}_cCQ^{j\gamma}_d)H^kH^l(H^\dag H)$ \\
 \hline
\end{longtable}
\end{center}

\vspace{-7ex}

\section{Dimension-11 operators}
\label{sec:dim-11-ops}

\subsection{\texorpdfstring{$(\Delta B, \Delta L) = (1, -1), \ov{\ell}^2\psi^4\varphi^2$}{1, -1, l2psi4H2}}
\label{subsec:1, -1, l2psi4H2}

\subsubsection{\texorpdfstring{$\ov{L}^2\psi^4\varphi^2$}{L2psi4H2}}
\label{subsubsec:L2psi4H2}

\begin{center}
\renewcommand{\arraystretch}{1.3}
\begin{longtable}{l l} 
 \hline
 Operator type & Terms \\ [0.5ex] 
 \hline\hline
 \multirow{2}{*}{$\wt{H}^2\ov{L}^2u^2de$} & $\eps_{\alpha\beta\gamma}(\ov{L}_ad^\alpha_b\wt{H})(\ov{L}_cu^\beta_d\wt{H})(e_eCu^\gamma_f)$ \\
  & $\eps_{\alpha\beta\gamma}(\ov{L}_ae_b\wt{H})(\ov{L}_cu^\alpha_d\wt{H})(u^\beta_eCd^\gamma_f)$ \\
 \hline
 \multirow{3}{*}{$\wt{H}^2\ov{L}^2udQL$} & $\eps_{\alpha\beta\gamma}\eps_{ij}(\ov{L}_ad^\alpha_b\wt{H})(\ov{L}_cu^\beta_d\wt{H})(L^i_eCQ^{j\gamma}_f)$ \\
  & $\eps_{\alpha\beta\gamma}\delta_{ik}\eps_{jl}(\ov{L}_ad^\alpha_b\wt{H})(\ov{L}^i_cu^\beta_d)(L^j_eCQ^{k\gamma}_f)\wt{H}^l$ \\
  & $\eps_{\alpha\beta\gamma}\eps_{ij}\eps_{km}\eps_{ln}(\ov{L}^i_ad^\alpha_b)(\ov{L}^j_cu^\beta_d)(L^k_eCQ^{l\gamma}_f)\wt{H}^m\wt{H}^n$ \\
 \hline
 \multirow{2}{*}{$\wt{H}^2\ov{L}^2uQ^2e$} & $\eps_{\alpha\beta\gamma}\delta_{ij}\eps_{kl}(\ov{L}_ae_b\wt{H})(\ov{L}^i_cu^\alpha_d)(Q^{j\beta}_eCQ^{k\gamma}_f)\wt{H}^l$ \\
  & $\eps_{\alpha\beta\gamma}\eps_{ij}\eps_{km}\eps_{ln}(\ov{L}^i_ae_b)(\ov{L}^j_cu^\alpha_d)(Q^{k\beta}_eCQ^{l\gamma}_f)\wt{H}^m\wt{H}^n$ \\
\hline
\end{longtable}
\end{center}

\subsubsection{\texorpdfstring{$\ov{L}\ov{Q}\psi^4\varphi^2$}{LQpsi4H2}}
\label{subsubsec:LQpsi4H2}

\begin{center}
\renewcommand{\arraystretch}{1.3}
\begin{longtable}{l l} 
 \hline
 Operator type & Terms \\ [0.5ex] 
 \hline\hline
 \multirow{4}{*}{$\wt{H}^2\ov{L}\ov{Q}u^2d^2$} & $\delta_{\alpha\beta}\epsilon_{\gamma\rho\sigma}(\ov{Q}_{a}^{\alpha}u_b^\beta\wt{H})(u_c^\gamma C d_d^\rho)(\ov{L}_e d_f^\sigma\wt{H})$ \\
  & $\delta_{\alpha\beta}\epsilon_{\gamma\rho\sigma}(\ov{Q}_{a}^{\alpha}d_b^\beta\wt{H})(u_c^\gamma C d_d^\rho)(\ov{L}_e u_f^\sigma\wt{H})$ \\
  & $\delta_{\alpha\sigma}\epsilon_{\beta\gamma\rho}(\ov{Q}_{a}^{\alpha}\sigma_{\mu\nu}u_b^\beta\wt{H})(u_c^\gamma C\sigma^{\mu\nu}d_d^\rho)(\ov{L}_e d_f^\sigma\wt{H})$ \\
  & $\delta_{\alpha\rho}\epsilon_{\beta\gamma\sigma}(\ov{Q}_{a}^{\alpha}\sigma_{\mu\nu}u_b^\beta\wt{H})(\ov{L}_c\sigma^{\mu\nu}d_d^\gamma\wt{H})(u_e^\rho C d_f^\sigma)$ \\
 \hline
 \multirow{10}{*}{$\wt{H}^2\ov{L}\ov{Q}udQ^2$} & $\delta_{\alpha\beta}\epsilon_{\gamma\rho\sigma}\delta_{ij}\epsilon_{kl}
 (\ov{Q}_a^{i\alpha}d_b^\beta)(\ov{L}_c u_d^\gamma\wt{H})
 (Q_e^{j\rho} C Q_f^{k\sigma})\wt{H}^l$ \\
 & $\delta_{\alpha\gamma}\epsilon_{\beta\rho\sigma}\delta_{ij}\epsilon_{kl}
 (\ov{Q}_a^{i\alpha}d_b^\beta)(\ov{L}_c u_d^\gamma\wt{H})
 (Q_e^{j\rho} C Q_f^{k\sigma})\wt{H}^l$ \\
 & $\delta_{\alpha\rho}\epsilon_{\beta\gamma\sigma}\delta_{ij}\epsilon_{kl}
 (\ov{Q}_a^{i\alpha}d_b^\beta)(\ov{L}_c u_d^\gamma\wt{H})
 (Q_e^{j\rho} C Q_f^{k\sigma})\wt{H}^l$ \\
\hline
\end{longtable}
\end{center}

\vspace{-4ex}

\subsection{\texorpdfstring{$(\Delta B, \Delta L) = (1, -1), \ov{\ell}\ov{r}\psi^4\varphi^2$}{1, -1, lrpsi4H2}}
\label{subsec:1, -1, lrpsi4H2}

\subsubsection{\texorpdfstring{$\ov{L}\ov{e}\psi^4\varphi^2$}{Lepsi4H2}}
\label{subsubsec:Lepsi4H2}

\begin{center}
\renewcommand{\arraystretch}{1.3}
\begin{longtable}{l l} 
 \hline
 Operator type & Terms \\ [0.5ex] 
 \hline\hline
 $\wt{H}^2\ov{L}\ov{e}ud^2L$ & $\eps_{\alpha\beta\gamma}(\ov{L}_ad^\alpha_b\wt{H})(H^\dag\ov{e}_cL_d)(d^\beta_eCu^\gamma_f)$ \\
 \hline
 \multirow{2}{*}{$\wt{H}^2\ov{L}\ov{e}udQe$} & $\eps_{\alpha\beta\gamma}(\ov{L}_ad^\alpha_b\wt{H})(H^\dag \ov{e}_cQ^\beta_d)(e_eCu^\gamma_f)$ \\
  & $\eps_{\alpha\beta\gamma}(\ov{L}_au^\alpha_b\wt{H})(H^\dag \ov{e}_cQ^\beta_d)(e_eCd^\gamma_f)$ \\
 \hline
 \multirow{3}{*}{$\wt{H}^2\ov{L}\ov{e}dQ^2L$} & $\eps_{\alpha\beta\gamma}\eps_{ij}(\ov{L}_ad^\alpha_b\wt{H})(H^\dag\ov{e}_cQ^\beta_d)(L^i_eCQ^{j\gamma}_f)$ \\
  & $\eps_{\alpha\beta\gamma}\delta_{ij}\eps_{kl}(\ov{L}^i_ad^\alpha_b)(H^\dag\ov{e}_cQ^\beta_d)(L^j_eCQ^{k\gamma}_f)\wt{H}^l$ \\
  & $\eps_{\alpha\beta\gamma}\eps_{ij}\eps_{kl}(\ov{L}_ad^\alpha_b\wt{H})(\ov{e}_cQ^{i\beta}_d)(L^j_eCQ^{k\gamma}_f)\wt{H}^l$ \\
 \hline
 $\wt{H}^2\ov{L}\ov{e}Q^3e$ & $\eps_{\alpha\beta\gamma}\delta_{ik}\eps_{jm}\eps_{ln}(\ov{L}^i_ae_b)(\ov{e}_cQ^{j\alpha}_d)(Q^{k\beta}_eCQ^{l\gamma}_f)\wt{H}^m\wt{H}^n$ \\
 \hline
 \multirow{2}{*}{$\wt{H}\ov{L}\ov{e}d^3LH$} & $\eps_{\alpha\beta\gamma}\delta_{ij}(\ov{L}^i_ad^\alpha_b)(\ov{e}_cL^j_d)(d^\beta_eCd^\gamma_f)(H^\dag H)$ \\
 & $\eps_{\alpha\beta\gamma}(\ov{L}_ad^\alpha_bH)(H^\dag\ov{e}_cL_d)(d^\beta_eCd^\gamma_f)$ \\
 \hline
 \multirow{2}{*}{$\wt{H}\ov{L}\ov{e}d^2QeH$} & $\eps_{\alpha\beta\gamma}(\ov{L}_ad^\alpha_bH)(H^\dag\ov{e}_cQ^\beta_d)(e_eCd^\gamma_f)$ \\
  & $\eps_{\alpha\beta\gamma}(\ov{L}_ad^\alpha_b\wt{H})(\wt{H}^\dag\ov{e}_cQ^\beta_d)(e_eCd^\gamma_f)$ \\
 \hline
\end{longtable}
\end{center}

\vspace{-8ex}

\subsubsection{\texorpdfstring{$\ov{L}\ov{d}\psi^4\varphi^2$}{Ldpsi4H2}}
\label{subsubsec:Ldpsi4H2}

\begin{center}
\renewcommand{\arraystretch}{1.3}
\begin{longtable}{l l} 
 \hline
 Operator type & Terms \\ [0.5ex] 
 \hline\hline
 \multirow{3}{*}{$\wt{H}^2\ov{L}\ov{d}ud^2Q$} & $\delta_{\alpha\sigma}\eps_{\beta\gamma\rho}(H^\dag\ov{d}^\alpha_aQ^\beta_b)(\ov{L}_cd^\gamma_d\wt{H})(d^\rho_eCu^\sigma_f)$ \\
  & $\delta_{\alpha\rho}\eps_{\beta\gamma\sigma}(H^\dag\ov{d}^\alpha_aQ^\beta_b)(\ov{L}_cd^\gamma_d\wt{H})(d^\rho_eCu^\sigma_f)$  \\
  & $\delta_{\alpha\gamma}\eps_{\beta\rho\sigma}(H^\dag\ov{d}^\alpha_aQ^\beta_b)(\ov{L}_cd^\gamma_d\wt{H})(d^\rho_eCu^\sigma_f)$  \\
 \hline
 \multirow{3}{*}{$\wt{H}^2\ov{L}\ov{d}dQ^3$} & $\delta_{\alpha\sigma}\eps_{\beta\gamma\rho}\delta_{ij}\eps_{kl}(H^\dag\ov{d}^\alpha_aQ^\beta_b)(\ov{L}^i_cd^\gamma_d)(Q^{j\rho}_eCQ^{k\sigma}_f)\wt{H}^l$ \\
  & $\delta_{\alpha\rho}\eps_{\beta\gamma\sigma}\delta_{ij}\eps_{kl}(H^\dag\ov{d}^\alpha_aQ^\beta_b)(\ov{L}^i_cd^\gamma_d)(Q^{j\rho}_eCQ^{k\sigma}_f)\wt{H}^l$ \\
  & $\delta_{\alpha\sigma}\eps_{\beta\gamma\rho}\eps_{ij}\eps_{kl}(\ov{d}^\alpha_aQ^{i\beta}_b)(\ov{L}_cd^\gamma_d\wt{H})(Q^{j\rho}_eCQ^{k\sigma}_f)\wt{H}^l$ \\
 \hline
 \multirow{2}{*}{$\wt{H}\ov{L}\ov{d}d^3QH$} & $\delta_{\alpha\sigma}\eps_{\beta\gamma\rho}(H^\dag\ov{d}^\alpha_aQ^\beta_b)(\ov{L}_cd^\gamma_dH)(d^\rho_eCd^\sigma_f)$ \\
  & $\delta_{\alpha\sigma}\eps_{\beta\gamma\rho}(\wt{H}^\dag\ov{d}^\alpha_aQ^\beta_b)(\ov{L}_cd^\gamma_d\wt{H})(d^\rho_eCd^\sigma_f)$ \\
 \hline
\end{longtable}
\end{center}

\subsubsection{\texorpdfstring{$\ov{L}\ov{u}\psi^4\varphi^2$}{Lupsi4H2}}
\label{subsubsec:Lupsi4H2}

\begin{center}
\renewcommand{\arraystretch}{1.3}
\begin{longtable}{l l} 
 \hline
 Operator type & Terms \\ [0.5ex] 
 \hline\hline
 \multirow{3}{*}{$\wt{H}^2\ov{L}\ov{u}u^2dQ$} & $\delta_{\alpha\sigma}\eps_{\beta\gamma\rho}(H^\dag\ov{u}^\alpha_aQ^\beta_b)(\ov{L}_cu^\gamma_d\wt{H})(u^\rho_eCd^\sigma_f)$ \\
  & $\delta_{\alpha\rho}\eps_{\beta\gamma\sigma}(H^\dag\ov{u}^\alpha_aQ^\beta_b)(\ov{L}_cu^\gamma_d\wt{H})(u^\rho_eCd^\sigma_f)$  \\
  & $\delta_{\alpha\gamma}\eps_{\beta\rho\sigma}(H^\dag\ov{u}^\alpha_aQ^\beta_b)(\ov{L}_cu^\gamma_d\wt{H})(u^\rho_eCd^\sigma_f)$  \\ \hline
 \multirow{3}{*}{$\wt{H}^2\ov{L}\ov{u}uQ^3$} & $\delta_{\alpha\sigma}\eps_{\beta\gamma\rho}\delta_{ij}\eps_{kl}(H^\dag\ov{u}^\alpha_aQ^\beta_b)(\ov{L}^i_cu^\gamma_d)(Q^{j\rho}_eCQ^{k\sigma}_f)\wt{H}^l$ \\
  & $\delta_{\alpha\rho}\eps_{\beta\gamma\sigma}\delta_{ij}\eps_{kl}(H^\dag\ov{u}^\alpha_aQ^\beta_b)(\ov{L}^i_cu^\gamma_d)(Q^{j\rho}_eCQ^{k\sigma}_f)\wt{H}^l$ \\
  & $\delta_{\alpha\sigma}\eps_{\beta\gamma\rho}\eps_{ij}\eps_{kl}(\ov{u}^\alpha_aQ^{i\beta}_b)(\ov{L}_cu^\gamma_d\wt{H})(Q^{j\rho}_eCQ^{k\sigma}_f)\wt{H}^l$ \\
 \hline
 \multirow{6}{*}{$\wt{H}\ov{L}\ov{u}ud^2QH$} & $\delta_{\alpha\sigma}\eps_{\beta\gamma\rho}(H^\dag\ov{u}^\alpha_aQ^\beta_b)(\ov{L}_cd^\gamma_dH)(d^\rho_eCu^\sigma_f)$ \\
\hline
\end{longtable}
\end{center}

\newpage

\subsubsection{\texorpdfstring{$\ov{Q}\ov{e}\psi^4\varphi^2$}{eQpsi4H2}}
\label{subsubsec:eQpsi4H2}

\begin{center}
\renewcommand{\arraystretch}{1.3}
\begin{longtable}{l l} 
 \hline
 Operator type & Terms \\ [0.5ex] 
 \hline\hline
 \multirow{3}{*}{$\wt{H}^2\ov{e}\ov{Q}ud^2Q$} & $\delta_{\alpha\sigma}\eps_{\beta\gamma\rho}(\ov{Q}^\alpha_ad^\beta_b\wt{H})(H^\dag\ov{e}_cQ^\gamma_d)(d^\rho_eCu^\sigma_f)$ \\
  & $\delta_{\alpha\rho}\eps_{\beta\gamma\sigma}(\ov{Q}^\alpha_ad^\beta_b\wt{H})(H^\dag\ov{e}_cQ^\gamma_d)(d^\rho_eCu^\sigma_f)$ \\
  & $\delta_{\alpha\gamma}\eps_{\beta\rho\sigma}(\ov{Q}^\alpha_ad^\beta_b\wt{H})(H^\dag\ov{e}_cQ^\gamma_d)(d^\rho_eCu^\sigma_f)$ \\
 \hline
 \multirow{3}{*}{$\wt{H}^2\ov{e}\ov{Q}dQ^3$} & $\delta_{\alpha\sigma}\eps_{\beta\gamma\rho}\delta_{ij}\eps_{kl}(\ov{Q}^{i\alpha}_ad^\beta_b)(H^\dag\ov{e}_cQ^{\gamma}_d)(Q^{j\rho}_eCQ^{k\sigma}_f)\wt{H}^l$ \\
  & $\delta_{\alpha\rho}\eps_{\beta\gamma\sigma}\delta_{ij}\eps_{kl}(\ov{Q}^{i\alpha}_ad^\beta_b)(H^\dag\ov{e}_cQ^{\gamma}_d)(Q^{j\rho}_eCQ^{k\sigma}_f)\wt{H}^l$ \\
  & $\delta_{\alpha\gamma}\eps_{\beta\rho\sigma}\delta_{ij}\eps_{kl}(\ov{Q}^{i\alpha}_ad^\beta_b)(H^\dag\ov{e}_cQ^{\gamma}_d)(Q^{j\rho}_eCQ^{k\sigma}_f)\wt{H}^l$ \\
 \hline
 \multirow{2}{*}{$\wt{H}\ov{e}\ov{Q}d^3QH$} & $\delta_{ij}\delta_{\alpha\sigma}\eps_{\beta\gamma\rho}(\ov{Q}^{i\alpha}_ad^\beta_b)(\ov{e}_cQ^{j\gamma}_d)(d^\rho_eCd^\sigma_f)(H^\dag H)$ \\
  & $\delta_{\alpha\sigma}\eps_{\beta\gamma\rho}(\ov{Q}^{\alpha}_ad^\beta_bH)(H^\dag\ov{e}_cQ^{\gamma}_d)(d^\rho_eCd^\sigma_f)$ \\
 \hline
\end{longtable}
\end{center}

\subsection{\texorpdfstring{$(\Delta B, \Delta L) = (1, -1), \ov{r}^2\psi^4\varphi^2$}{1, -1, r2psi4H2}}
\label{subsec:1, -1, r2psi4H2}

\begin{center}
\renewcommand{\arraystretch}{1.3}
\begin{longtable}{l l} 
 \hline
 Operator type & Terms \\ [0.5ex] 
 \hline\hline
 $\wt{H}^2\ov{e}^2d^2QL$ & $\eps_{\alpha\beta\gamma}(H^\dag\ov{e}_aQ^{\alpha}_b)(H^\dag\ov{e}_cL_d)(d^\beta_eCd^\gamma_f)$ \\
 $\wt{H}^2\ov{e}^2dQ^2e$ & $\eps_{\alpha\beta\gamma}(H^\dag\ov{e}_aQ^\alpha_b)(H^\dag\ov{e}_cQ^\beta_d)(e_eCd^\gamma_f)$ \\
 $\wt{H}\ov{e}^2d^3eH$ & $\eps_{\alpha\beta\gamma}(\ov{e}_ae^\C_b)(e_cCd^\alpha_d)(d^\beta_eCd^\gamma_f)(H^\dag H)$ \\
 \hline
 \multirow{2}{*}{$\wt{H}^2\ov{e}\ov{d}d^2Q^2$} & $\delta_{\alpha\sigma}\eps_{\beta\gamma\rho}(H^\dag\ov{d}^\alpha_aQ^\beta_b)(H^\dag\ov{e}_cQ^\gamma_d)(d^\rho_eCd^\sigma_f)$ \\
  & $\delta_{\alpha\gamma}\eps_{\beta\rho\sigma}(H^\dag\ov{d}^\alpha_aQ^\beta_b)(H^\dag\ov{e}_cQ^\gamma_d)(d^\rho_eCd^\sigma_f)$ \\
 \hline
 $\wt{H}\ov{e}\ov{d}d^4H$ & $\delta_{\alpha\sigma}\eps_{\beta\gamma\rho}(\ov{d}^{\alpha}_ae^\C_b)(d^\beta_cCd^\gamma_d)(d^\rho_eCd^\sigma_f)(H^\dag H)$ \\
 \hline
 \multirow{3}{*}{$\wt{H}^2\ov{e}\ov{u}udQ^2$} & $\delta_{\alpha\sigma}\eps_{\beta\gamma\rho}(H^\dag\ov{u}^\alpha_aQ^\beta_b)(H^\dag\ov{e}_cQ^\gamma_d)(d^\rho_eCu^\sigma_f)$ \\
\hline
\end{longtable}
\end{center}

\clearpage

\subsection{\texorpdfstring{$(\Delta B, \Delta L) = (1, -1), \ov{\psi}\psi^3\varphi^5$}{1, -1, psi4H5}}
\label{subsec:1, -1, psi4H5}

\begin{center}
\renewcommand{\arraystretch}{1.3}
\begin{longtable}{l l} 
 \hline
 Operator type & Terms \\ [0.5ex] 
 \hline\hline
 $\wt{H}^3\ov{L}ud^2H^2$ & $\eps_{\alpha\beta\gamma}(\ov{L}_ad^\alpha_b\wt{H})(d^\beta_cCu^\gamma_d)(H^\dag H)(H^\dag H)$ \\
 \hline
 \multirow{2}{*}{$\wt{H}^3\ov{L}dQ^2H^2$} & $\eps_{\alpha\beta\gamma}\eps_{ik}\eps_{jl}(\ov{L}_ad^\alpha_b\wt{H})(Q^{i\beta}_cCQ^{j\gamma}_d)\wt{H}^kH^l(H^\dag H)$ \\
  & $\eps_{\alpha\beta\gamma}\eps_{ik}\eps_{jl}(\ov{L}_ad^\alpha_bH)(Q^{i\beta}_cCQ^{j\gamma}_d)\wt{H}^k\wt{H}^l(H^\dag H)$ \\
 \hline
 $\wt{H}^3\ov{e}d^2QH^2$ & $\eps_{\alpha\beta\gamma}(\ov{e}_aQ^\alpha_b\wt{H})(d^\beta_cCd^\gamma_d)(H^\dag H)(H^\dag H)$ \\
 $\wt{H}^2\ov{L}d^3H^3$ & $\eps_{\alpha\beta\gamma}(\ov{L}_ad^\alpha_bH)(d^\beta_cCd^\gamma_d)(H^\dag H)(H^\dag H)$ \\
 $\wt{H}^4\ov{e}Q^3H$ & $\eps_{\alpha\beta\gamma}\eps_{ik}\eps_{jl}(H^\dag\ov{e}_aQ^\alpha_b)(Q^{i\beta}_cCQ^{j\gamma}_d)\wt{H}^k\wt{H}^l(H^\dag H)$ \\
 $\wt{H}^4\ov{L}uQ^2H$ & $\eps_{\alpha\beta\gamma}\eps_{ik}\eps_{jl}(\ov{L}_au^\alpha_b\wt{H})(Q^{i\beta}_cCQ^{j\gamma}_d)\wt{H}^k\wt{H}^l(H^\dag H)$ \\
 \hline
\end{longtable}
\end{center}

\subsection{\texorpdfstring{$(\Delta B, \Delta L) = (1, 3), \psi^6\varphi^2$}{1, 3, psi6H2}}
\label{subsec:1, 3, psi6H2}

\begin{center}
\renewcommand{\arraystretch}{1.3}
\begin{longtable}{l l} 
 \hline
 Operator type & Terms \\ [0.5ex] 
 \hline\hline
 $\wt{H}u^3eL^2H$ & $\eps_{\alpha\beta\gamma}\eps_{ik}\eps_{jl}(L^i_aCL^j_b)(e_cCu^\alpha_d)(u^\beta_eCu^\gamma_f)\wt{H}^kH^l$ \\
 \hline
 \multirow{2}{*}{$\wt{H}u^2QL^3H$} & $\eps_{\alpha\beta\gamma}\eps_{il}\eps_{jm}\eps_{kn}(Q^{i\alpha}_aCL^j_b)(L^k_cCL^l_d)(u^\beta_eCu^\gamma_f)\wt{H}^mH^n$ \\
  & $\eps_{\alpha\beta\gamma}\eps_{il}\eps_{jn}\eps_{km}(Q^{i\alpha}_aCL^j_b)(L^k_cCL^l_d)(u^\beta_eCu^\gamma_f)\wt{H}^mH^n$ \\
 \hline
 $u^2deL^2H^2$ & $\eps_{\alpha\beta\gamma}\eps_{ik}\eps_{jl}(L^i_aCL^j_b)(e_cCu^\alpha_d)(u^\beta_eCd^\gamma_f)H^kH^l$ \\
 $u^2Qe^2LH^2$ & $\eps_{\alpha\beta\gamma}\eps_{ik}\eps_{jl}(L^i_aCQ^{j\alpha}_b)(e_cCu^\beta_d)(e_eCu^\gamma_f)H^kH^l$ \\
 \hline
 $udQL^3H^2$ & $\eps_{\alpha\beta\gamma}\eps_{il}\eps_{jm}\eps_{kn}(Q^{i\alpha}_aCL^j_b)(L^k_cCL^l_d)(d^\beta_eCu^\gamma_f)H^mH^n$ \\
 \hline
 \multirow{2}{*}{$uQ^2eL^2H^2$} & $\eps_{\alpha\beta\gamma}\eps_{il}\eps_{jm}\eps_{kn}(L^i_aCQ^{j\alpha}_b)(L^k_cCQ^{l\beta}_d)(e_eCu^\gamma_f)H^mH^n$ \\
  & $\eps_{\alpha\beta\gamma}\eps_{ij}\eps_{km}\eps_{ln}(L^i_aCQ^{j\alpha}_b)(L^k_cCQ^{l\beta}_d)(e_eCu^\gamma_f)H^mH^n$ \\
 \hline
 \multirow{2}{*}{$Q^3L^3H^2$} & $\eps_{\alpha\beta\gamma}\eps_{il}\eps_{jp}\eps_{kq}\eps_{mn}(L^i_aCQ^{j\alpha}_b)(L^k_cCQ^{l\beta}_d)(L^m_eCQ^{n\gamma}_f)H^pH^q$ \\
  & $\eps_{\alpha\beta\gamma}\eps_{il}\eps_{jp}\eps_{km}\eps_{nq}(L^i_aCL^j_b)(L^k_cCQ^{l\alpha}_d)(Q^{m\beta}_eCQ^{n\gamma}_f)H^pH^q$ \\
 \hline
\end{longtable}
\end{center}

\vspace{-4ex}

\subsection{\texorpdfstring{$(\Delta B, \Delta L) = (2, 0), \psi^6\varphi^2$}{2, 0, psi6H2}}
\label{subsec:2, 0, psi6H2}

\begin{center}
\renewcommand{\arraystretch}{1.3}
\begin{longtable}{l l} 
 \hline
 Operator type & Terms \\ [0.5ex] 
 \hline\hline
 \multirow{3}{*}{$\wt{H}^2u^2d^2Q^2$} & $\eps_{\alpha\gamma\sigma}\eps_{\beta\rho\tau}\eps_{ik}\eps_{jl}(Q^{i\alpha}_aCQ^{j\beta}_b)(d^\gamma_cCd^\rho_d)(u^\sigma_eCu^\tau_f)\wt{H}^k\wt{H}^l$ \\
  & $\eps_{\alpha\gamma\tau}\eps_{\beta\rho\sigma}\eps_{ik}\eps_{jl}(Q^{i\alpha}_aCQ^{j\beta}_b)(d^\gamma_cCu^\rho_d)(d^\sigma_eCu^\tau_f)\wt{H}^k\wt{H}^l$ \\
  & $\eps_{\alpha\beta\gamma}\eps_{\rho\sigma\tau}\eps_{ik}\eps_{jl}(Q^{i\alpha}_aCQ^{j\beta}_b)(d^\gamma_cCu^\rho_d)(d^\sigma_eCu^\tau_f)\wt{H}^k\wt{H}^l$ \\
 \hline
 \multirow{2}{*}{$\wt{H}^2udQ^4$} & $\eps_{\alpha\gamma\sigma}\eps_{\beta\rho\tau}\eps_{il}\eps_{jm}\eps_{kn}(Q^{i\alpha}_aCQ^{j\beta}_b)(Q^{k\gamma}_cCQ^{l\rho}_d)(d^\sigma_eCu^\tau_f)\wt{H}^m\wt{H}^n$ \\
  & $\eps_{\alpha\beta\sigma}\eps_{\gamma\rho\tau}\eps_{il}\eps_{jm}\eps_{kn}(Q^{i\alpha}_aCQ^{j\beta}_b)(Q^{k\gamma}_cCQ^{l\rho}_d)(d^\sigma_eCu^\tau_f)\wt{H}^m\wt{H}^n$ \\
 \hline
 $\wt{H}^2Q^6$ & $\eps_{\alpha\gamma\sigma}\eps_{\beta\rho\tau}\eps_{ip}\eps_{jk}(\eps\tau^I)_{lm}(\eps\tau^I)_{nq}(Q^{i\alpha}_aCQ^{j\beta}_b)(Q^{k\gamma}_cCQ^{l\rho}_d)(Q^{m\sigma}_eCQ^{n\tau}_f)\wt{H}^p\wt{H}^q$ \\
 \hline
 \multirow{2}{*}{$\wt{H}u^2d^4H$} & $\eps_{\alpha\gamma\sigma}\eps_{\beta\rho\tau}(d^\alpha_aCu^\beta_b)(d^\gamma_cCu^\rho_d)(d^\sigma_eCd^\tau_f)(H^\dag H)$ \\
\hline
\end{longtable}
\end{center}

\vspace{-5ex}

\section{Symmetry-protected dimension-12 operators}
\label{sec:dim-12-ops}

\subsection{\texorpdfstring{$(\Delta B, \Delta L) = (2, -2), \ov{\psi}^2\psi^6$}{2, -2, psi8}}
\label{subsec:2, -2, psi8}

\begin{center}
\renewcommand{\arraystretch}{1.3}
\begin{longtable}{l l} 
 \hline
 Operator type & Terms \\ [0.5ex] 
 \hline\hline
 $\ov{L}^2ud^5$ & $\eps_{\alpha\gamma\sigma}\eps_{\beta\rho\tau}\eps_{ij}(\ov{L}^i_ad^\alpha_b)(\ov{L}^j_cd^\beta_d)(d^\gamma_eCd^\rho_f)(d^\sigma_gCu^\tau_h)$ \\
 \hline
 \multirow{2}{*}{$\ov{L}^2d^4Q^2$} & $\eps_{\alpha\gamma\sigma}\eps_{\beta\rho\tau}\delta_{ik}\delta_{jl}(\ov{L}^i_ad^\alpha_b)(\ov{L}^j_cd^\beta_d)(d^\gamma_eCd^\rho_f)(Q^{k\sigma}_gCQ^{l\tau}_h)$ \\
  & $\eps_{\alpha\gamma\sigma}\eps_{\beta\rho\tau}\eps_{ij}\eps_{kl}(\ov{L}^i_ad^\alpha_b)(\ov{L}^j_cd^\beta_d)(d^\gamma_eCd^\rho_f)(Q^{k\sigma}_gCQ^{l\tau}_h)$ \\
 \hline
 $\ov{L}\ov{e}d^5Q$ & $\eps_{\alpha\gamma\sigma}\eps_{\beta\rho\tau}\delta_{ij}(\ov{L}^i_ad^\alpha_b)(\ov{e}_cQ^{j\beta}_d)(d^\gamma_eCd^\rho_f)(d^\sigma_gCd^\tau_h)$ \\
 $\ov{e}^2d^6$ & $\eps_{\alpha\gamma\sigma}\eps_{\beta\rho\tau}(\ov{e}_ae^\C_b)(d^\alpha_cCd^\beta_d)(d^\gamma_eCd^\rho_f)(d^\sigma_gCd^\tau_h)$ \\
 \hline
\end{longtable}
\end{center}

\vspace{-6ex}

\subsection{\texorpdfstring{$(\Delta B, \Delta L) = (2, 2), \psi^8$}{2, 2, psi8}}
\label{subsec:2, 2, psi8}

\begin{center}
\renewcommand{\arraystretch}{1.3}
\begin{longtable}{l l} 
 \hline
 Operator type & Terms \\ [0.5ex] 
 \hline\hline
 \multirow{2}{*}{$u^4d^2e^2$} & $\eps_{\alpha\rho\sigma}\eps_{\beta\gamma\tau}(d^\alpha_aCu^\beta_b)(d^\gamma_cCu^\rho_d)(e_eCu^\sigma_f)(e_gCu^\tau_h)$ \\
  & $\eps_{\alpha\beta\sigma}\eps_{\gamma\rho\tau}(d^\alpha_aCe_b)(d^\beta_cCu^\gamma_d)(e_eCu^\rho_f)(u^\sigma_gCu^\tau_h)$ \\
 \hline
 \multirow{2}{*}{$u^3d^3L^2$} & $\eps_{\alpha\gamma\sigma}\eps_{\beta\rho\tau}\eps_{ij}(d^\alpha_aCd^\beta_b)(d^\gamma_cCu^\rho_d)(u^\sigma_eCu^\tau_f)(L^i_gCL^j_h)$ \\
  & $\eps_{\alpha\gamma\tau}\eps_{\beta\rho\sigma}\eps_{ij}(d^\alpha_aCu^\beta_b)(d^\gamma_cCu^\rho_d)(d^\sigma_eCu^\tau_f)(L^i_gCL^j_h)$ \\
 \hline
 \multirow{3}{*}{$u^3d^2QeL$} & $\eps_{\alpha\rho\sigma}\eps_{\beta\gamma\tau}\eps_{ij}(d^\alpha_aCu^\beta_b)(d^\gamma_cCu^\rho_d)(e_eCu^\sigma_f)(L^i_gCQ^{j\tau}_h)$ \\
  & $\eps_{\alpha\beta\rho}\eps_{\gamma\sigma\tau}\eps_{ij}(d^\alpha_aCe_b)(d^\beta_cCu^\gamma_d)(u^\rho_eCu^\sigma_f)(L^i_gCQ^{j\tau}_h)$ \\
  & $\eps_{\alpha\gamma\sigma}\eps_{\beta\rho\tau}\eps_{ij}(d^\alpha_aCd^\beta_b)(e_cCu^\gamma_d)(u^\rho_eCu^\sigma_f)(L^i_gCQ^{j\tau}_h)$ \\
\hline
\end{longtable}
\end{center}

\section{Symmetry-protected dimension-14 operators}
\label{sec:dim-14-ops}

\subsection{\texorpdfstring{$(\Delta B, \Delta L) = (1, 5), \psi^8\phi^2$}{1, 5, psi8H2}}
\label{subsec:1, 5, psi8H2}

\begin{center}
\renewcommand{\arraystretch}{1.3}
\begin{longtable}{l l} 
 \hline
 Operator type & Terms \\ [0.5ex] 
 \hline\hline
 $u^3 e L^4 H^2$ & $\epsilon_{\alpha\beta\gamma}\epsilon_{il}\epsilon_{jm}\epsilon_{kn}(L_a^i C L_b^j)(L_c^k C L_d^l)(e_e C u_f^\alpha)(u_g^\beta C u_h^\gamma)H^m H^n$ \\
 $u^2 Q L^5 H^2$ & $\epsilon_{\alpha\beta\gamma}\epsilon_{im}\epsilon_{jp}\epsilon_{kn}\epsilon_{lq}(Q_a^{i\alpha} C L_b^j)(L_c^k C L_d^l)(L_e^m C L_f^n)(u_g^\beta C u_h^\gamma)H^p H^q$ \\
 \hline
\end{longtable}
\end{center}

\section{Symmetry-protected dimension-15 operators}
\label{sec:dim-15-ops}

\subsection{\texorpdfstring{$(\Delta B, \Delta L) = (1, -5), \ov\psi^5\psi^3\phi^3$}{1, -5, psi*5psi3H3}}
\label{subsec:1, -5, psi*5psi3H3}

\begin{center}
\renewcommand{\arraystretch}{1.3}
\begin{longtable}{l l} 
 \hline
 Operator type & Terms \\ [0.5ex] 
 \hline\hline
 $\wt H^3\ov L^5 d^3$ & $\epsilon_{\alpha\beta\gamma}\delta_{il}\epsilon_{jk}(\ov L_a^i L_b^{\C,j})(\ov L_c^k d_d^\alpha)(\ov L_e d_f^\beta\widetilde H)(\ov L_g d_h^\gamma\widetilde H)\widetilde H^l$ \\
 \hline
\end{longtable}
\end{center}

\vspace{-3ex}

\subsection{\texorpdfstring{$(\Delta B, \Delta L) = (2, -4), \ov\psi^4\psi^6$}{2, -4, psi*4psi6}}
\label{subsec:2, -4, psi*4psi6}

\begin{center}
\renewcommand{\arraystretch}{1.3}
\begin{longtable}{l l} 
 \hline
 Operator type & Terms \\ [0.5ex] 
 \hline\hline
 $\ov L^4d^6$ & $\epsilon_{\alpha\gamma\tau}\epsilon_{\beta\rho\sigma}\epsilon_{ij}\epsilon_{kl}(\ov L_a^i d_b^\alpha)(\ov L_c^j d_d^\beta)(\ov L_e^k d_f^\gamma)(\ov L_g^l d_h^\rho)(d_r^\sigma C d_s^\tau)$ \\
 \hline
\end{longtable}
\end{center}

\vspace{-3ex}

\subsection{\texorpdfstring{$(\Delta B, \Delta L) = (2, 4), \psi^{10}$}{2, 4, psi10}}
\label{subsec:2, 4, psi10}

\begin{center}
\renewcommand{\arraystretch}{1.3}
\begin{longtable}{l l} 
 \hline
 Operator type & Terms \\ [0.5ex] 
 \hline\hline
 $u^5 Q e^3 L$ & $\epsilon_{\alpha\beta\sigma}\epsilon_{\gamma\rho\tau}\epsilon_{ij}(L_a^{i} C Q_b^{j\alpha})(e_c C u_d^{\beta})(e_e C u_f^{\gamma})(e_g C u_h^{\rho})(u_r^{\sigma} C u_s^{\tau})$ \\
 $u^5 d e^2 L^2$ & $\epsilon_{\alpha\rho\sigma}\epsilon_{\beta\gamma\tau}\epsilon_{ij}(L_a^{i} C L_b^{j})(d_c^{\alpha} C u_d^{\beta})(e_e C u_f^{\gamma})(e_g C u_h^{\rho})(u_r^{\sigma} C u_s^{\tau})$ \\
 \hline
 \multirow{2}{*}{$u^4 Q^2 e^2 L^2$} & $\epsilon_{\alpha\gamma\sigma}\epsilon_{\beta\rho\tau}\epsilon_{ij}\epsilon_{kl}(L_a^{i} C Q_b^{j\alpha})(L_c^{k} C Q_d^{l\beta})(e_e C u_f^{\gamma})(e_g C u_h^{\rho})(u_r^{\sigma} C u_s^{\tau})$ \\
  & $\epsilon_{\alpha\gamma\sigma}\epsilon_{\beta\rho\tau}\epsilon_{ik}\epsilon_{jl}(L_a^{i} C Q_b^{j\alpha})(L_c^{k} C Q_d^{l\beta})(e_e C u_f^{\gamma})(e_g C u_h^{\rho})(u_r^{\sigma} C u_s^{\tau})$ \\
 \hline
 \multirow{2}{*}{$u^4 d Q e L^3$} & $\epsilon_{\alpha\rho\sigma}\epsilon_{\beta\gamma\tau}\epsilon_{ik}\epsilon_{jl}(L_a^{i} C L_b^{j})(L_c^{k} C Q_d^{l\alpha})(d_e^{\beta} C u_f^{\gamma})(e_g C u_h^{\rho})(u_r^{\sigma} C u_s^{\tau})$ \\
  & $\epsilon_{\alpha\gamma\sigma}\epsilon_{\beta\rho\tau}\epsilon_{ik}\epsilon_{jl}(L_a^{i} C L_b^{j})(L_c^{k} C Q_d^{l\alpha})(d_e^{\beta} C u_f^{\gamma})(e_g C u_h^{\rho})(u_r^{\sigma} C u_s^{\tau})$ \\
 \hline
 \multirow{2}{*}{$u^3 Q^3 e L^3$} & $\epsilon_{\alpha\beta\sigma}\epsilon_{\gamma\rho\tau}\epsilon_{im}\epsilon_{jn}\epsilon_{kl}(L_a^{i} C Q_b^{j\alpha})(L_c^{k} C Q_d^{l\beta})(L_e^{m} C Q_f^{n\gamma})(e_g C u_h^{\rho})(u_r^{\sigma} C u_s^{\tau})$ \\
\hline
\end{longtable}
\end{center}

The numbers of terms shown for $u^3dQ^2L^4$ and $u^2Q^4L^4$ rely on the fact that there are three fermion generations. If the four $L$ flavor labels are treated formally, the minimum numbers of terms $N_{\mathcal T}$ are two and three, respectively. The larger counts arise from combinations that are totally antisymmetric in the four $L$ flavor labels, which vanish when those labels can only take three values. The minimum numbers of terms are then reduced to one and two, respectively, as given in our tables.

\vspace{-0.5ex}

\subsection{\texorpdfstring{$(\Delta B, \Delta L) = (3, 1), \psi^{10}$}{3, 1, psi10}}
\label{subsec:3, 1, psi10}

\begin{center}
\renewcommand{\arraystretch}{1.3}
\begin{longtable}{l l} 
 \hline
 Operator type & Terms \\ [0.5ex] 
 \hline\hline
 $u^3 d^5 Q L$ & $\epsilon_{\alpha\kappa\lambda}\epsilon_{\beta\rho\sigma}\epsilon_{\gamma\delta\tau}\epsilon_{ij}(L_a^{i} C Q_b^{j\alpha})(d_c^{\beta} C d_d^{\gamma})(d_e^{\delta} C u_f^{\kappa})(d_g^{\lambda} C u_h^{\rho})(d_r^{\sigma} C u_s^{\tau})$ \\
 \hline
 \multirow{3}{*}{$u^3 d^4 Q^2 e$} & $\epsilon_{\alpha\gamma\kappa}\epsilon_{\beta\lambda\rho}\epsilon_{\delta\sigma\tau}\epsilon_{ij}(Q_a^{i\alpha} C Q_b^{j\beta})(d_c^{\gamma} C e_d)(d_e^{\delta} C u_f^{\kappa})(d_g^{\lambda} C u_h^{\rho})(d_r^{\sigma} C u_s^{\tau})$ \\
  & $\epsilon_{\alpha\gamma\lambda}\epsilon_{\beta\rho\sigma}\epsilon_{\delta\kappa\tau}\epsilon_{ij}(Q_a^{i\alpha} C Q_b^{j\beta})(d_c^{\gamma} C e_d)(d_e^{\delta} C u_f^{\kappa})(d_g^{\lambda} C u_h^{\rho})(d_r^{\sigma} C u_s^{\tau})$ \\
  & $\epsilon_{\alpha\gamma\lambda}\epsilon_{\beta\rho\sigma}\epsilon_{\delta\kappa\tau}\epsilon_{ij}(Q_a^{i\alpha} C Q_b^{j\beta})(d_c^{\gamma} C d_d^{\delta})(d_e^{\kappa} C u_f^{\lambda})(d_g^{\rho} C u_h^{\sigma})(e_r C u_s^{\tau})$ \\
 \hline
 $u^4 d^5 e$ & $\epsilon_{\alpha\gamma\lambda}\epsilon_{\beta\rho\sigma}\epsilon_{\delta\kappa\tau}(d_a^{\alpha} C e_b)(d_c^{\beta} C u_d^{\gamma})(d_e^{\delta} C u_f^{\kappa})(d_g^{\lambda} C u_h^{\rho})(d_r^{\sigma} C u_s^{\tau})$ \\
 \hline
 \multirow{4}{*}{$u^2 d^4 Q^3 L$} & $\epsilon_{\alpha\lambda\rho}\epsilon_{\beta\delta\sigma}\epsilon_{\gamma\kappa\tau}\epsilon_{ik}\epsilon_{jl}(L_a^{i} C Q_b^{j\alpha})(Q_c^{k\beta} C Q_d^{l\gamma})(d_e^{\delta} C d_f^{\kappa})(d_g^{\lambda} C u_h^{\rho})(d_r^{\sigma} C u_s^{\tau})$ \\
  & $\epsilon_{\alpha\rho\sigma}\epsilon_{\beta\delta\lambda}\epsilon_{\gamma\kappa\tau}\epsilon_{ik}\epsilon_{jl}(L_a^{i} C Q_b^{j\alpha})(Q_c^{k\beta} C Q_d^{l\gamma})(d_e^{\delta} C d_f^{\kappa})(d_g^{\lambda} C u_h^{\rho})(d_r^{\sigma} C u_s^{\tau})$ \\
  & $\epsilon_{\alpha\delta\rho}\epsilon_{\beta\gamma\kappa}\epsilon_{\lambda\sigma\tau}\epsilon_{ik}\epsilon_{jl}(L_a^{i} C Q_b^{j\alpha})(Q_c^{k\beta} C Q_d^{l\gamma})(d_e^{\delta} C d_f^{\kappa})(d_g^{\lambda} C u_h^{\rho})(d_r^{\sigma} C u_s^{\tau})$ \\
  & $\epsilon_{\alpha\gamma\kappa}\epsilon_{\beta\lambda\rho}\epsilon_{\delta\sigma\tau}\epsilon_{ij}\epsilon_{kl}(L_a^{i} C Q_b^{j\alpha})(Q_c^{k\beta} C Q_d^{l\gamma})(d_e^{\delta} C d_f^{\kappa})(d_g^{\lambda} C u_h^{\rho})(d_r^{\sigma} C u_s^{\tau})$ \\
 \hline
 \multirow{2}{*}{$u d^2 Q^6 e$} & $\epsilon_{\alpha\kappa\rho}\epsilon_{\beta\delta\sigma}\epsilon_{\gamma\lambda\tau}\epsilon_{ik}\epsilon_{jm}\epsilon_{ln}(Q_a^{i\alpha} C Q_b^{j\beta})(Q_c^{k\gamma} C Q_d^{l\delta})(Q_e^{m\kappa} C Q_f^{n\lambda})(d_g^{\rho} C e_h)(d_r^{\sigma} C u_s^{\tau})$ \\
  & $\epsilon_{\alpha\kappa\rho}\epsilon_{\beta\gamma\sigma}\epsilon_{\delta\lambda\tau}\epsilon_{im}\epsilon_{jn}\epsilon_{kl}(Q_a^{i\alpha} C Q_b^{j\beta})(Q_c^{k\gamma} C Q_d^{l\delta})(Q_e^{m\kappa} C Q_f^{n\lambda})(d_g^{\rho} C e_h)(d_r^{\sigma} C u_s^{\tau})$ \\
 \hline
 \multirow{5}{*}{$u^2 d^3 Q^4 e$} & $\epsilon_{\alpha\delta\sigma}\epsilon_{\beta\lambda\tau}\epsilon_{\gamma\kappa\rho}\epsilon_{ij}\epsilon_{kl}(Q_a^{i\alpha} C Q_b^{j\beta})(Q_c^{k\gamma} C Q_d^{l\delta})(d_e^{\kappa} C e_f)(d_g^{\lambda} C u_h^{\rho})(d_r^{\sigma} C u_s^{\tau})$ \\
  & $\epsilon_{\alpha\beta\rho}\epsilon_{\gamma\kappa\sigma}\epsilon_{\delta\lambda\tau}\epsilon_{ik}\epsilon_{jl}(Q_a^{i\alpha} C Q_b^{j\beta})(Q_c^{k\gamma} C Q_d^{l\delta})(d_e^{\kappa} C d_f^{\lambda})(d_g^{\rho} C u_h^{\sigma})(e_r C u_s^{\tau})$ \\
  & $\epsilon_{\alpha\delta\lambda}\epsilon_{\beta\sigma\tau}\epsilon_{\gamma\kappa\rho}\epsilon_{ik}\epsilon_{jl}(Q_a^{i\alpha} C Q_b^{j\beta})(Q_c^{k\gamma} C Q_d^{l\delta})(d_e^{\kappa} C e_f)(d_g^{\lambda} C u_h^{\rho})(d_r^{\sigma} C u_s^{\tau})$ \\
  & $\epsilon_{\alpha\beta\lambda}\epsilon_{\gamma\kappa\sigma}\epsilon_{\delta\rho\tau}\epsilon_{ik}\epsilon_{jl}(Q_a^{i\alpha} C Q_b^{j\beta})(Q_c^{k\gamma} C Q_d^{l\delta})(d_e^{\kappa} C d_f^{\lambda})(d_g^{\rho} C u_h^{\sigma})(e_r C u_s^{\tau})$ \\
  & $\epsilon_{\alpha\delta\lambda}\epsilon_{\beta\rho\tau}\epsilon_{\gamma\kappa\sigma}\epsilon_{ij}\epsilon_{kl}(Q_a^{i\alpha} C Q_b^{j\beta})(Q_c^{k\gamma} C Q_d^{l\delta})(d_e^{\kappa} C d_f^{\lambda})(d_g^{\rho} C u_h^{\sigma})(e_r C u_s^{\tau})$ \\
 \hline
 \multirow{4}{*}{$u d^3 Q^5 L$} & $\epsilon_{\alpha\lambda\tau}\epsilon_{\beta\gamma\sigma}\epsilon_{\delta\kappa\rho}\epsilon_{im}\epsilon_{jk}\epsilon_{ln}(L_a^{i} C Q_b^{j\alpha})(Q_c^{k\beta} C Q_d^{l\gamma})(Q_e^{m\delta} C Q_f^{n\kappa})(d_g^{\lambda} C d_h^{\rho})(d_r^{\sigma} C u_s^{\tau})$ \\
  & $\epsilon_{\alpha\kappa\rho}\epsilon_{\beta\gamma\sigma}\epsilon_{\delta\lambda\tau}\epsilon_{ij}\epsilon_{km}\epsilon_{ln}(L_a^{i} C Q_b^{j\alpha})(Q_c^{k\beta} C Q_d^{l\gamma})(Q_e^{m\delta} C Q_f^{n\kappa})(d_g^{\lambda} C d_h^{\rho})(d_r^{\sigma} C u_s^{\tau})$ \\
  & $\epsilon_{\alpha\kappa\sigma}\epsilon_{\beta\gamma\rho}\epsilon_{\delta\lambda\tau}\epsilon_{ij}\epsilon_{km}\epsilon_{ln}(L_a^{i} C Q_b^{j\alpha})(Q_c^{k\beta} C Q_d^{l\gamma})(Q_e^{m\delta} C Q_f^{n\kappa})(d_g^{\lambda} C d_h^{\rho})(d_r^{\sigma} C u_s^{\tau})$ \\
  & $\epsilon_{\alpha\gamma\sigma}\epsilon_{\beta\kappa\rho}\epsilon_{\delta\lambda\tau}\epsilon_{ik}\epsilon_{jm}\epsilon_{ln}(L_a^{i} C Q_b^{j\alpha})(Q_c^{k\beta} C Q_d^{l\gamma})(Q_e^{m\delta} C Q_f^{n\kappa})(d_g^{\lambda} C d_h^{\rho})(d_r^{\sigma} C u_s^{\tau})$ \\
 \hline
 \multirow{2}{*}{$d^2 Q^7 L$} & $\epsilon_{\alpha\kappa\tau}\epsilon_{\beta\gamma\rho}\epsilon_{\delta\lambda\sigma}\epsilon_{ik}\epsilon_{jp}\epsilon_{lm}\epsilon_{nq}$ \\
  & $\qquad\times (L_a^{i} C Q_b^{j\alpha})(Q_c^{k\beta} C Q_d^{l\gamma})(Q_e^{m\delta} C Q_f^{n\kappa})(Q_g^{p\lambda} C Q_h^{q\rho})(d_r^{\sigma} C d_s^{\tau})$ \\
 \hline
 \multirow{2}{*}{$d Q^8 e$} & $\epsilon_{\alpha\beta\lambda}\epsilon_{\gamma\delta\sigma}\epsilon_{\kappa\rho\tau}\epsilon_{ik}\epsilon_{jp}\epsilon_{lm}\epsilon_{nq}$ \\
  & $\qquad\times (Q_a^{i\alpha} C Q_b^{j\beta})(Q_c^{k\gamma} C Q_d^{l\delta})(Q_e^{m\kappa} C Q_f^{n\lambda})(Q_g^{p\rho} C Q_h^{q\sigma})(d_r^{\tau} C e_s)$ \\
 \hline
\end{longtable}
\end{center}

\section{Conclusions}
\label{sec:conclusions}

In this work, we have constructed explicit bases for non-derivative baryon-number-violating SMEFT operators through dimension 11, together with the symmetry-protected non-derivative operators from dimensions 12 to 15. We have tried to present complete bases in a form that is as transparent and usable as possible. Specifically, whenever possible, we give bases that are both minimal and built from simple contraction patterns, with the few exceptions collected separately in appendix~\ref{app:nonminimal}. In this way, the results provide a concrete and organized description of a large portion of the BNV SMEFT landscape. The corresponding derivative operators, whose structure is substantially more involved, will be treated separately in a future paper~\cite{Heeck:DerivativeBNV}.

A central point of the paper is that minimality at the level of terms is a subtle problem. To address it, we formulated the basis problem in the flavor-permutation space, where the relevant linear relations can be studied cleanly, and combined three ingredients: a group-theoretic determination of the target operator-space dimension, a permutation-theoretic count of the minimal number of terms, and an explicit constructive verification that the chosen terms span the full space. This framework makes it possible to assess, in a systematic way, whether a proposed set of representatives is not only complete, but also minimal.

Our results also clarify the status of the existing literature. At low dimensions, we reproduce known minimal bases where they are correct, while at higher dimensions, we often obtain equivalent but simpler representatives, and in some cases show that previously proposed bases are not minimal. More broadly, the classification presented here shows how rapidly the number of BNV operator types grows with mass dimension, underscoring the need for explicit and well-organized bases if one wants to study this sector systematically.

The motivation for doing so is clear. Because baryon number is an accidental symmetry of the SM, its violation would be an unambiguous signal of new physics. At the same time, the extreme sensitivity of nucleon-decay and related searches means that operators well above dimension 6 can remain phenomenologically relevant. A reliable operator basis is therefore an essential prerequisite for exploring the full baryon-number-violating SMEFT landscape, especially at high mass dimension where many potentially relevant operators remain comparatively unstudied, and for connecting that broader space of possibilities to concrete observables and ultraviolet completions. We hope that the bases constructed here will provide such a foundation and help make the broader baryon-number-violating SMEFT program more systematic and accessible.

\section*{Acknowledgements}

J.H.~thanks Bryan Wright for assistance with \texttt{AutoEFT}; the \texttt{AutoEFT} basis for non-derivative BNV SMEFT operators up to $d=15$ can be found in Ref.~\cite{autoEFTbasis}.
This work was supported by the U.S.~Department of Energy under Grant No.~DE-SC0007974.
We acknowledge Research Computing at The University of Virginia for providing computational resources that have contributed to the results reported within this publication.

\appendix

\section{Nonminimal operator bases}
\label{app:nonminimal}

Almost all of the bases presented in sections~\ref{sec:dim-6-ops}--\ref{sec:dim-15-ops} are minimal. In a small number of cases, however, we choose to present slightly nonminimal bases instead. The reason is not that minimal bases fail to exist, but that the corresponding minimal representatives can be quite unnatural and significantly more cumbersome than nonminimal alternatives. Since one of our main goals is to give bases that are not only valid but also simple and usable, in such cases we regard a nice nonminimal basis as preferable to an ugly minimal one. The operator types for which the basis shown in the main text is not minimal are collected in table~\ref{tab:nonminimal-bases}.

\setcounter{table}{4}
\begin{table}[b]
    \centering
    \renewcommand{\arraystretch}{1.2}
    \begin{tabular}{lcc}
         Operator type & Our \#terms & Minimal basis \#terms \\ [0.5ex] 
         \hline
         $\wt{H}\ov{L}\ov{e}d^3LH$ (\ref{subsubsec:Lepsi4H2}) & 2 & 1 \\
         $\wt{H}^2u^2d^2Q^2$ (\ref{subsec:2, 0, psi6H2}) & 3 & 2 \\
         $u^2d^2Q^2L^2$ (\ref{subsec:2, 2, psi8}) & 5 & 4
    \end{tabular}
    \caption{Operator types for which we do not present a minimal basis in the main text.}
    \label{tab:nonminimal-bases}
\end{table}

In these cases, there is a structural reason why a nice minimal basis cannot be constructed, even though the minimal number of terms is known. We illustrate this point with the example $\wt{H}\ov{L}\ov{e}d^3LH$. Here, the obstruction comes from the interplay between flavor permutations and the symmetry properties forced by the gauge and Lorentz contractions. As a result, while a minimal basis exists in principle, any simple choice of term has too small a flavor-permutation orbit to span the full space, so one must either use an awkward minimal representative or enlarge the basis slightly. In other cases, a similar counting argument demonstrates that no nice minimal basis exists.

Consider the operator type $\T = \wt{H}\ov{L}\ov{e}d^3LH$, for which we provide a nonminimal basis in subsubsection~\ref{subsubsec:Lepsi4H2}. For this type, a minimal basis should contain only 1 term, while $\dim W_{\wt{H}\ov{L}\ov{e}d^3LH} = 4$. Since the only repeated field is $d$, the flavor-permutation group is $G_{\wt{H}\ov{L}\ov{e}d^3LH} = S_3$. Thus, if a single term $\O_{\wt{H}\ov{L}\ov{e}d^3LH}^{(1)}$ were sufficiently generic, one might hope that its six flavor-permuted operators $\qty{\mathcal O_{\und{\wt{H}\ov{L}\ov{e}d^3LH}{\pi\cdot\mathfrak F}}^{(1)}:\pi\in S_3}$ would contain at least four linearly independent operators and hence span the full space $W_{\wt{H}\ov{L}\ov{e}d^3LH}$. In other words, the group-theoretic counting by itself does not rule out a one-term minimal basis.

However, this maximal orbit size cannot be realized by any nice term due to the structure of the fields. The three $d$ fields are the only colored triplets in the operator, so their color indices must be contracted together with a single $\eps_{\alpha\beta\gamma}$. At the same time, in any simple Lorentz contraction pattern, two of the three \(d\)'s necessarily appear together inside one Lorentz bilinear after Fierz transformations, schematically as a factor of the form $(d C d)$, with the remaining $d$ contracted elsewhere. For example, simple representatives naturally have schematic structures such as $(\ov{L}d)(\ov{e}L)(dCd)\wt{H}H$ or $(\ov{L}L)(\ov{e}d)(dCd)\wt{H}H$, up to the detailed placement of gauge contractions. In each case, one distinguished pair of $d$-slots is grouped together.

Once two $d$'s are locked into the same bilinear, exchanging them cannot produce a new linearly independent operator. Indeed, under the transposition of those two $d$-slots, the term returns to itself up to a sign coming from the symmetry properties of the Lorentz and color contractions. This transposition therefore fixes the one-dimensional subspace spanned by the term. The stabilizer of the subspace has order at least 2, so the orbit-stabilizer theorem gives at most $6/2=3$ distinct subspaces in its $S_3$ orbit. Their span has dimension at most 3, below $\dim W_{\mathcal T} = 4$. 

This can also be seen more concretely. Suppose the paired $d$-fields are $d_a$ and $d_b$. Then $\O_{abc \cdots}^{(1)}$ and $\O_{bac \cdots}^{(1)}$ differ only by exchanging those two $d$'s within the same Lorentz bilinear $(dCd)$. Moreover, these same two fields also belong to the same color contraction, since all three $d$'s are contracted through a single $\eps_{\alpha\beta\gamma}$. As a result, exchanging $d_a$ and $d_b$ only changes the expression by the symmetry factors associated with the Lorentz and color contractions, so the two operators are proportional. The only genuinely different possibilities are therefore the three choices of which flavor label is singled out as the $d$ not belonging to that distinguished pair. Hence, a ``nice'' one-term representative can generate at most three independent operators.

Therefore, for any nice term, $\dim W_{\wt{H}\ov{L}\ov{e}d^3LH}^{(1)}\le 3$, whereas the full space has dimension $\dim W_{\wt{H}\ov{L}\ov{e}d^3LH} = 4$. This inequality shows that no nice one-term basis can span \(W_{\wt{H}\ov{L}\ov{e}d^3LH}\), even though the abstract minimal term count is~1. In this sense, a nice minimal basis is impossible for this operator type. One must therefore choose between an awkward minimal representative such as
\begin{equation}
\mathcal{O}^{(1)}_{\wt{H}\ov{L}\ov{e}d^3LH} = \eps_{\alpha\beta\gamma}\qty[(\ov{L}d^\alpha H)(H^\dag\ov{e}L)(d^\beta C d^\gamma) + \delta_{ij}(\ov{L}^i d^\beta)(\ov{e}L^j)(d^\alpha C d^\gamma)(H^\dag H)]
\end{equation}
and a slightly larger but much simpler nonminimal basis; in the main text, we adopt the latter.

 \bibliographystyle{JHEP}
 \bibliography{bib.bib}
 
\end{document}